\documentclass[conference]{IEEEtran} 
\IEEEoverridecommandlockouts

\usepackage{cite}
\usepackage[fleqn]{amsmath}
\usepackage{amsfonts,amssymb}
\usepackage{tabularx}
\usepackage{textcomp}
\usepackage{xcolor}
\usepackage{footnote}
\usepackage{float}
\usepackage[export]{adjustbox}
\usepackage{caption}
\usepackage{times}  
\usepackage{helvet} 
\usepackage{courier}  
\usepackage[hyphens]{url}  
\usepackage{graphicx} 
\usepackage{graphicx}  
\usepackage{multirow}
\usepackage{booktabs}

\usepackage{xcolor}
\usepackage{bm}

\usepackage{amsmath}
\usepackage{verbatim}
\usepackage{subcaption}
\usepackage{algorithm}
\usepackage{algpseudocode}

\makeatletter

\makeatletter
\newcommand{\linebreakand}{%
  \end{@IEEEauthorhalign}
  \hfill\mbox{}\par
  \mbox{}\hfill\begin{@IEEEauthorhalign}
}

\makeatother
\def\BibTeX{{\rm B\kern-.05em{\sc i\kern-.025em b}\kern-.08em
    T\kern-.1667em\lower.7ex\hbox{E}\kern-.125emX}}
\begin{document}

\title{Joint position and trajectory optimization of flying base station in 5G cellular networks, based on users' current and predicted location }

\author{\IEEEauthorblockN{Mehdi Sookhak and Amir Hossein Mohajerzadeh}

 \thanks{Mehdi Sookhak and Amir Hossein Mohajerzadeh with the Department of Computer Science and Engineering,Texas A\&M University at Galveston. 
The corresponding author is Mehdi Sookhak (email: m.sookhak@ieee.org)\newline
 This work has been submitted to the IEEE for possible publication.  Copyright may be transferred without notice, after which this version may no longer be accessible
 } 

 }
\maketitle

\begin{abstract}
Nowadays, Unmanned Aerial Vehicles (UAVs) have been significantly improved, and one of their most important applications is to provide temporary coverage for cellular users. Static Base Station cannot service all users due to temporary crashes because of temporary events such as ground BS breakdowns, bad weather conditions, natural disasters, transmission errors, etc., drones equipped with small cellular BS. The Drone Base Station is immediately sent to the target location and establishes the necessary communication links without requiring any predetermined infrastructure and covers that area. Finding the optimal location and the appropriate number (DBS) of drone-BS in this area is a challenge. Therefore, in this paper, the optimal location and optimal number of DBSs are distributed in the current state of the users and the subsequent user states determined by the prediction. Finally, the DBS transition is optimized from the current state to the predicted future locations. The simulation results show that the proposed method can provide acceptable coverage on the network.
\\
\end{abstract} 


\section{Introduction} \label{sec.1}
\label{sec:intro}
Wireless network users expect to have an unlimited access network with minimal cost per location and time. It is to be added that in mobile networks, user desires require high reliability and availability \cite{Li2021}. In cases where user density increases temporarily, QoS must still be guaranteed. A promising solution is a use of more ground stations (BSs), which are not affordable due to high OPEX and CAPEX costs. On the one hand, because of continuous changes in users' spatial and temporal densities, and the volume and the rate of data which users redirect, Traffic pattern prediction is difficult \cite{9586045}. Therefore, agility and resilience in cellular networks may not be met with the help of ground BS. Researchers recently reviewed the use of drones equipped with the drone BS small cellular BS module (DBS)\cite{Mozaffari2019}. The drone is an unmanned aerial vehicle that can be operated either remotely or autonomously with the help of software and sensors embedded in it. The main uses of drones are for military purposes to track and identify targets, but recently lightweight, battery-powered drones have been designed for civilian applications and areas with immediate and urgent needs. Drones often use a battery to feed the rotor and onboard electronic components\cite{8903295}. In mobile cellular networks, the biggest advantage of drones is that they are equipped with a small cellular BS, in remote and low-population areas, and in cases where there is a transmission error due to temporary events such as groundbreaking BS breakdowns or due to bad weather conditions, natural disasters, Subversion and sudden congestion in places such as sports stadiums, terrestrial BSs cannot serve all users due to overcrowding (as shown in Figure \ref{fig.1}) \cite{7962642}, they are immediately sent to the target location and provide essential communication links. Without the need for any predetermined infrastructure. As a result, DBS can move in space to improve the coverage and capacity of the network and largely provide QoS for the needs of cellular network users (even cell-edge users). There are challenges and limitations in using DBS as follows:
\begin{figure}[H] 
    \centering
    \includegraphics[width =0.95\columnwidth] {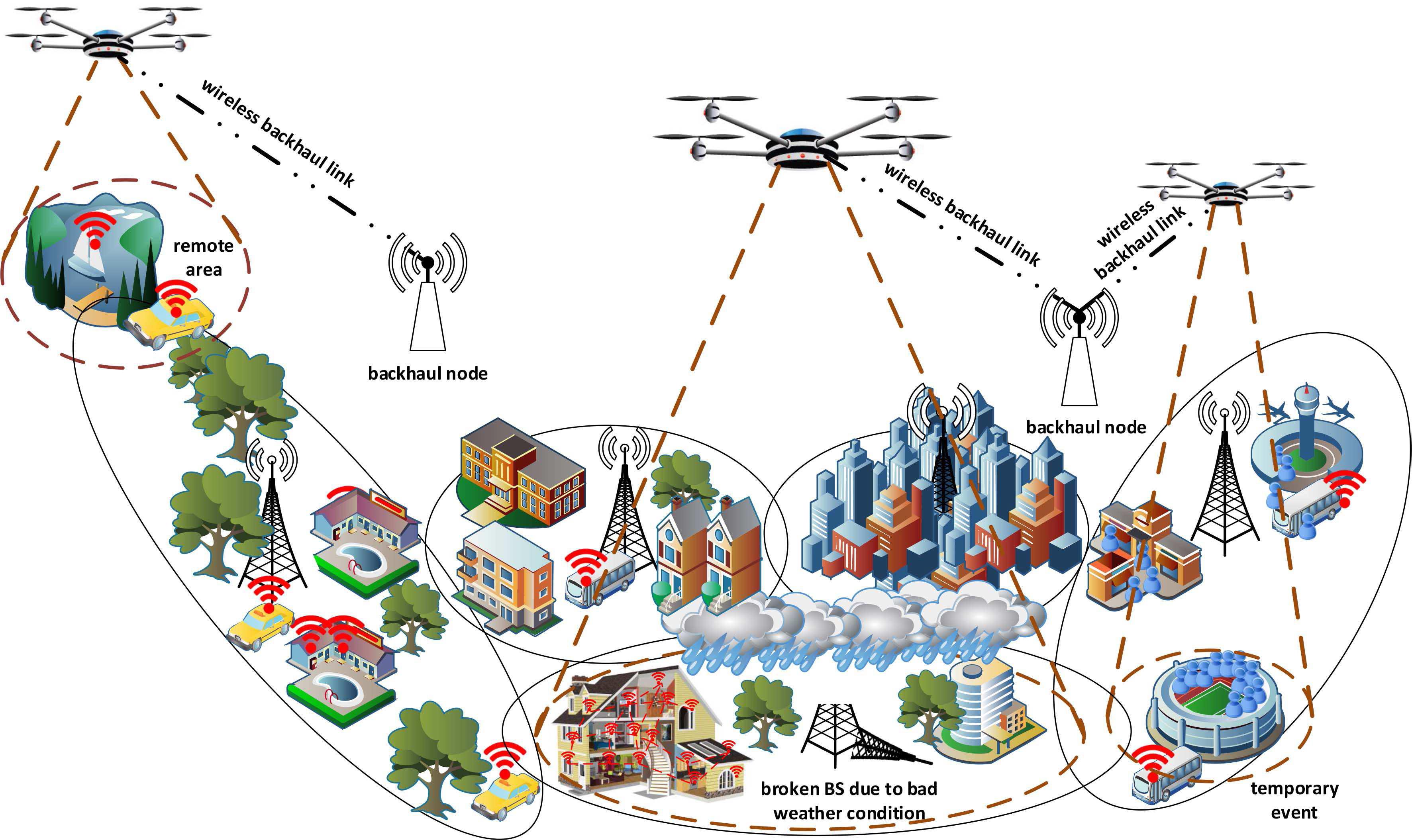}
    \caption{A possible scenario of temporary coverage of users by DBS}
    \label{fig.1}
\end{figure}
\subsection*{\textit{\textbf{ Height}}} 
DBS cannot fly at very high altitudes. If DBS fly at high altitudes, they should be made heavier for resistance and protection against winds, which will cause a high weight during a crash. Also, because of security reasons and regulatory constraints, such as pedestrian traffic and vehicles, DBS cannot fly at a very low altitude \cite{Fotouhi2019}, DBS should be located at an altitude where the number of covered users is maximized\cite{Lai2019}.
\subsection*{\textit{\textbf{ Speed}}} 
Due to hardware constraints, DBS cannot fly very fast; after a certain speed level of, say, 100 milliseconds, their energy consumption will increase, so they should not fly too fast\cite{Fotouhi2019}.
\subsection*{\textit{\textbf{ Agility}}} 
DBS is not very agile due to hardware constraints. Operation such as rotation or full stop will take a lot of time \cite{8760267,Fotouhi2019}
\subsection*{\textit{\textbf{ Other challenges are as follows}}} 
• If the DBS is not properly designed, its OPEX cost may be higher than the BS; so, you need to be careful in designing it to save on the overall cost of deploying the network \cite{Muntaha2021}.\par
• If the mission time and lifetime of the DBS battery become longer, its transmission capacity and hence the covered area will decrease, so there should be a tradeoff between transfer power, DBS mission, and tradeoff area.\par
Out of the Challenges Due to the importance of the impact of the DBS deployment on air-to-ground airline reliability, we also address this issue.
\section{Research Objectives}
\begin{itemize}
    \item Finding the optimal place and number of DBS for the current user status so that the maximum number of users is covered, in Section \ref{sec.4.1}, is examined in more detail.
    \item Currently, we have the BS number at the moment t with the number of active users, by predicting the number of users of each BS at the moment t + 1, and re-establishing the optimal number and location of the DBS with the goal of minimizing Path loss users. This issue is fully described in Section \ref{sec.4.2}.
    \item In Section \ref{sec.4.3}, we plan to optimize the DBS transfer from the current state to the next situation. Let's review the transition optimization between the present moment and the next moment in Section \ref{sec.4.3}.
\end{itemize}
\subsection*{\textit{\textbf{ Continue this article as follows}}} 
Section \ref{sec.2} of the recent research in this area (Related Work and Motivation) is introduced. In Section \ref{sec.3}, the system model describes the network and channel  models. Section \ref{sec.4} issues and solutions are raised. In Section \ref{sec.5}, the results of the theory and method of deployment of DBS are validated with numerical results. Finally, Section \ref{sec.6} concludes.
\section{Related Works}\label{sec.2}
Recently, thanks to resilience and agility, drones are used in wireless sensor networks and cellular networks. The issue of deploying DBS has been largely influenced by the reliability of the air-to-ground link, which is the main purpose of optimizing the location, height and number of DBS, in order to maximize coverage on the network. In this section, we examine DBS research into two categories.
\subsection{Optimal Drone Deployment}
The main difference between the terrestrial BS and DBS is that DBS has a lot of restrictions on the backhaul link. A static base station usually has a fixed wire link and can provide a high data rate for a Core network, while the DBS must have a wireless backlink. The wireless backhaul is an In-band Backhaul and Dedicated Backhaul. In a dedicated bachelor, there may be an FSO link or mmWAVE link between the Core network and the access network. These links provide high backhaul capacity, but are very sensitive to weather conditions, and in peak weather and rainy weather, peak rates are reduced \cite{9378782,8755983}. On the other hand, the current technology is the main backhaul links in LTE, Wi-Fi, HSPA and Microwave RF \cite{8771139} networks, which can be deployed very quickly and at a low cost. In the use of RF for backhaul, a spectrum is used in both access links and the backhaul link so interference is created and as a result of the capacity of the backhaul link is affected and the number of users in the service is severely restricted. Therefore, \cite{9423546} a method is proposed that considers the constraints and requirements of the backhaul link system in the design and implementation of DBS in 5G + networks. In this method, an optimal algorithm for deploying DBS aware of the backhaul is provided, which provides a centralized solution for finding the best 3D location of a DBS, assuming that the general view of the network is available in a central controller, and the number of users to be covered and served by user centric and network centric. In the network centric mode, users are selected regardless of network requirements, so most users with less data rates are selected, while users are selected on a user-centric basis, based on their priority. 
\subsection{User Priority Identification Criteria}
\subsection*{\textit{\textbf{ Sum-Rate}}} 
If the user's maximum sum-rate is the maximum and has higher requirements, then the higher priority will be given to accessing the network. Price Differentiation: Users are prioritized and categorized based on the cost they pay for subscribers, such as platinum, gold and silver users, who pay platinum users more money and expect that in every situation, even if the channel  is poor or more resources need to be connected to the network.
\subsection*{\textit{\textbf{ Signal Strength}}} 
Users are selected based on their received signal strength, meaning users who have better channel  conditions and are prioritized to the rest. Content Demand: In a content aware system, users who need immediate access are given a higher priority based on their content. After selecting the DBS location and coverage area, the robustness of the DBS is checked. \cite{9515712} A three-dimensional optimal UAV-BS deployment method is proposed for a variety of users in NOMA Networks, which by maximizing the transmission power, maximizes the number of users covered (coverage area). The goal of this article is boost the period of a communication coverage. For simplicity, the authors categorize the users and appropriates the transmission power of the all FBSs. In this paper, power consumption in BS is considered as a measure of energy efficiency used by the UAV-BS. Larger battery life and longer mission life will reduce UAV-BS transmission power, which will reduce UAV-BS coverage and thus reduce the number of users being served. As a result, between the transfer power, the UAV-BS and the area covered and the time of the trade-off mission.\par
In \cite{9296771}, a proactive drone-cell deployment framework has been proposed to reduce the overload caused by flash crowd traffic in 5G networks. Assumes the problem of deploying drone-cells as a clustering problem that considers the set of users assigned to the drone-cell as a cluster. The establishment of a drone-cell in the center of the cluster ensures that the minimum summation of the square of the distance is up to all cluster members. So finally, a constraint bisecting k-means is proposed to solve the drone-cell deployment problem. It also examines traffic patterns for 3 social activities: stadiums, parades, and gathering.
In \cite{9373692}, an active deployment method for cache-enabled drones is proposed to improve the quality of experience of file sharing system. An efficient algorithm proposed to improved network performance. This algorithm improves caching process, system throughput, and file sharing latency in V2X networks.
In \cite{9086619}, writers put forward the optimal 3D location of UAVs that are equipped with directional antennas using the circle packing concept, so that the overall coverage of the area is maximized. 
In \cite{Lai2019} an analytical model for finding the optimal height of an UAV with the aim of maximizing the coverage of the area is provided.
\cite{9044857} They have been investigated in finding optimal cell boundaries and deployment locations for multiple non interferential UAVs. The purpose of this paper is to minimize the power of the entire UAV. 
In \cite{9293315} writers transform the non-convex problem into an efficient power allocation problem for UAV 3D placement.
\subsection{Mobile DBSs}
The goal of \cite{9177323} is joint optimization of power consumption and hover time of UAVs under the quality of service constraint. The solution is achieved under this non-linear constrain by the Lagrange multipliers method and implemented the sub-gradient method. 
\cite{8320772} the autonomous motion control algorithm for DBS is designed and, by determining the best direction of movement for each drone, using the Game Theory Based Strategy, a Signal to Leakage Ratio (SLR) and SNR-based (Signal to Noise Ratio) strategies improve spectral efficiency of the network. In fact, the proposed system model is intended to be a network of multiple cells, each of which has a predetermined number of cells and a DBS in it, which can be moved at a constant speed and in a high elevation. Also, users of multiple mobile users are moving around each cell. Drone is connected to the nearby macro-cell tower through a wireless backhaul link that provides wireless communications for cellular users. As the drone moves, the distance between the users and the drone changes, and the drone moves closer to users, the signal strength gets improved. So, the direction of the drone must be constantly updated to suit the situation that users request. Moving each drone causes interference to active users within neighboring cells. Therefore, a SLR-based strategy is used that not only reduces the signal received to its active users but also interacts with other active users of neighboring cells. So the SLR is calculated by the drone and the drone moves in the direction of the largest SLR. Each drone focuses on its active users and assuming that there are no other neighboring drones, it calculates the SNR according to the location of its active users and it moves in the direction in which the maximum SNR is average, hence it improves the low-density drone spectral efficiency of the low-altitude. 
\cite{9293315} propose a method for maximization of the sum-rate in mmW networks. To aim this goal, they jointly optimized UAV location in 3D space, beam pattern, and transmit power. these sub-problems are solved in sequence. Firstly, standard convex optimization is used to find drone 2D locations. Then, the beam pattern is optimized by the multiobjective evolutionary algorithm. Finally, the 3D location of the UAV is derived. 
\cite{Zhang2021} jointly optimized 3D placement of UAV, user association, and frequency allocation attempt to increase coverage rate and the number of required UAVs under the QoS and UAV's serviceability constraints. The problem is formulated as a mixed-integer problem and it is solved in three steps. First, maximum user coverage is obtained by optimizing the required transmission power. Then, the minimum required drone is calculated with the help of an artificial bee colony (ABC) algorithm. Finally,  3D position and the frequency band are calculated to reduce interference along with increasing the power of the target signals.
swarm intelligence technology (SIT) is an algorithm witch work heuristically and used a random population as an evolutionary unit. The SIT is used in many industrial problems. \cite{Chen2020} used SIT based on the synergistic evolution between individuals of the population to optimize UAV deployment in IoT Data Acquisition scenario. 

While UAVs' potential mobility offers exciting possibilities, they also introduce new technological hurdles. UAV pathways must be optimized regarding the essential performance parameters such as throughput, energy efficiency, and latency \cite{Hong2021,Zhang12021}. In addition, route optimization challenges for the dynamic characteristics and types of drones should be described. 1) UAV route optimization to optimize communication throughput and UAV energy consumption, 2) Joint optimization of multi-user communication timing and UAV route to minimize latency, 3) Optimization Track optimization to maximize reliability in UAV-enabled wireless networks, and 4) Joint communications control and UAV route optimization to minimize flight time are all open problems in UAV route optimization\cite{Mozaffari2019}.
Numerous prior research excluded obstacles to ease the task of determining the optimal path. The communication link between the UAV and the users is also presumed to be line of sight in this case, however, a few papers have addressed the NLoS link \cite{Zhang2018,Hu2018}.

In \cite{Hua2020}, the UAV chain movement is controlled to improve the system throughput  between many moving and stable nodes. The 3D UAV trajectory, communication scheduling, and UAV-AP/SN transmit power are optimized.The polyblock outer approximation (POA) method is used to obtain a globally optimum solution. The problem of optimizing the path of constant power allocation is given as a non-convex optimization problem. Matching the location of a UAV serves as a relay to examine data from mobile users and forward them to the next base station \cite{Hiraguri2020}. In this work, it is assumed that the UAV can predict the user's location using any prediction algorithm. The goal is to optimize the users' accessibility uplink.
Additionally, there are numerous articles in previous works that discuss path loss as a crucial issue in the quality of service in determining the optimal path \cite{Wu2018,Ji2020,Tang2021,Zhao2019}. 

\subsection*{\textit{\textbf{ Contribution}}}
\begin{enumerate}
  \item Offering an optimization model taking into account the input and output parameters.
  \item Offering solution to solve the nonlinear optimization problem.
  \item Modeling the movement of users in the target environment.
  \item Determining the optimal mapping between DBS position at time t and t + 1.
  \item Determining the effectiveness of simultaneous DBS numbers and their locations.
  \item Use real data.
\end{enumerate}
The issue of deployment of DBS is of great importance and is largely influenced by the energy consumption as well as the interference produced by DBS. However, only a limited number of available studies point to the relationship between DBS deployment and network performance, and do not consider deployment with the minimum number of DBSs. The drawback of the DBS deployment methods is that a larger number of DBSs are needed to cover All users provide. Therefore, we first formulate the issue with Section \ref{sec.4.1} in order to maximize the number of users covered, as well as minimize the users of Path loss that are not covered by the current state of the users. In Section \ref{sec.4.2}, the future status of users of each BS is predicted, and we recalculate the optimal DBS number and location for maximum user coverage, as in Section \ref{sec.4.1}. Finally, in Section \ref{sec.4.3}, we will optimize the transfer of DBS from the current location to the next predicted location.
\section{System Model}\label{sec.3}
In this section, we examine the network and channel  models used in the proposed method.
\subsection{Network Model}
\begin{figure}[H] 
    \centering
    \includegraphics[width =0.95\columnwidth] {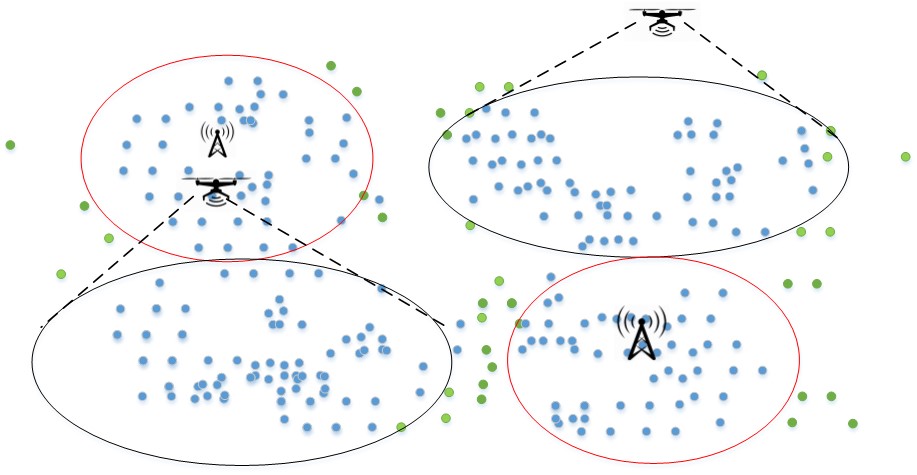}
    \caption{A possible scenario of user coverage by BS and DBS}
    \label{fig.2}
\end{figure}

Consider a region where a number of users have been broadcast. A stationary base station set $BS_i$, $i={1,2,...,NSBS}$ with specified coordinates $(x_B, y_B)$ has been deployed to serve terrestrial users in the environment. We assume that the number of users covered by each BS $(BS_i)$ is known at different times and is shown as $UBS_i$ $i \ \epsilon \ \{1,2,...,NU\}$. Suddenly, users of BS terrestrial devices cannot serve all of the users due to temporary overcrowding due to temporary events such as groundbreaking BS breakdowns or due to bad weather conditions, natural disasters, transmission errors, vandalism and congestion. Some DBSs are used to provide temporary coverage in these areas. Suppose we have a DBS set with coordinates $(x_d, y_d)$ in the form $DBS_i =(xd_i,yd_i) i \ \epsilon \ \{1,2,...,ndb\} $ to provide the full coverage of users in the region. Finding the optimal DBS location and minimum number of them is a problem that we will discuss in this article. We assume that the initial DBS location is in areas where user density is greater. Figure \ref{fig.2} illustrates a possible state of the DBS deployment issue that Macrocell has been overwhelmed, areas where users cannot be served by BSs are shown in red.

\subsection{Channel Model}

\begin{equation} \label{eq.1}{
\mathop{P(Los)=\dfrac{1}{1 +\alpha exp(-b(\dfrac{180}{\pi}\theta - \alpha))}}}
\end{equation}
\begin{equation}\label{eq.2}
\mathop{\theta = \arctan \dfrac{h}{r}} 
\end{equation}
According to equation \eqref{eq.1}, the probability of having a LoS relationship increases with increasing elevation angle, and if the horizontal distance $r$ is constant, the LoS probability increases with increasing the height of the harbor $h$. 
\begin{equation}\label{eq.3}
\mathop{P(LoS) = p(NLos)} 
\end{equation}
Equation \eqref{eq.3} shows $\eta LoS $ and $\eta NLoS $ are the mean of additional losses for $LoS$ and $NLoS$.
\begin{equation} \label{eq.4}
\begin{split}
pathloss & = 20  log(\dfrac{4\pi f_cd}{c})  + P(Los)\eta LoS \\
 & + P(NLoS)\eta NLoS
\end{split}
\end{equation}
In \eqref{eq.4} $f_c$ carrier frequency, c speed of light, d is the distance between the drone and the receiver, which is equivalent with $\sqrt{h^2 + r^2} $. 
\section{Problem statements and solutions}\label{sec.4}
\begin{figure}[H] 
    \centering
    \includegraphics[width =0.95\columnwidth] {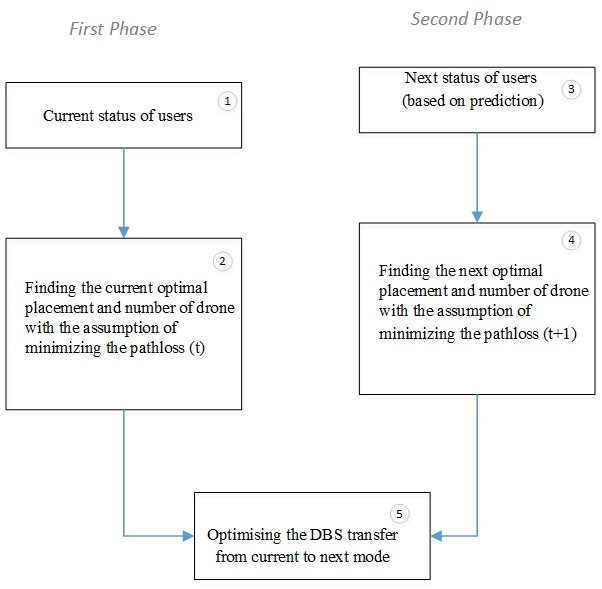}
    \caption{Overview of the proposed method}
    \label{fig.3}
\end{figure}

Figure \ref{fig.3} shows the general outline of the proposed method and the steps involved.
\subsection{Find the optimal place and number of DBS}\label{sec.4.1}
Given the current status of users that can be used by the actual university data, we calculate the optimal number and location of the DBS. We model this issue as follows:
\begin{equation}\label{eq.5}
\mathop{ Maximum_{xD,yD} \dfrac{\sum_{i=1}^{NDBS}\sum_{j=1}^{NU} UCD[i,j]}{totalUsers}}
\end{equation}

\begin{equation} \label{eq.6}
\begin{split}
& \forall i \in \{1,...,NDBS\}  \\
 & \sum_{j=1}^{NU}UCD_{i,j}*U_{bw} < DBS_{Maxbw}
\end{split}
\end{equation}
\begin{equation}\label{eq.7}
\mathop{pathloss \leq PL_{th}}
\end{equation}

\begin{equation}\label{eq.8}
\mathop{ \sum_{i=1}^{NDBS} DBS \leq 1 }
\end{equation}

In equation \eqref{eq.5} we want to maximize the proportion of total users of the DBS service to the total number of users. $NU$ shows the total number of network users and NDBS DBS numbers. Users have different rates for different applications. The overall bandwidth available to each DBS ($DBS_{Maxbw}$ Mbit/s) is limited to the limitations. The backhaul bet (condition \eqref{eq.6}) shows the bandwidth used by the user. In other words, to maintain the coverage of the entire user, the total bandwidth used by users of each DBS should be lower than the maximum bandwidth of DBS, so that all users will be served. $UCD_{i,j}$  is a user index function. $UCD_{I,j} \in \{0,1\}$ is binary variable that indicates whether the $j$ user is served with DBS. So, the $UCD{i,j}$ variable is 1 if the user $j$ is covered by my DBS, otherwise it is 0. Condition \eqref{eq.7} states that the path loss should be less than the threshold throttling threshold. We actually try to keep the path loss amount at least. In the channel  model (section \ref{sec.3}), the details of the path loss allocation are explained. Condition \eqref{eq.8} states that each user can have at least a subset and is covered by a DBS.\par
Because the problem is nonlinear, the optimization problem cannot be solved with existing methods. Therefore, we use a proposed algorithm to solve this nonlinear problem. In fact, we use the proposed algorithm to find the final solution ($\beta^*$) to the $p$ $Max_D$ roblem. Note that the maximum value of in the first run of the algorithm can be $Max_D$ (maximum number of DBS), under no circumstances can the DBS number be greater than $Max_D$; and minimum value of $\beta^*$ is 1 , since at least one DBS is required for the coating. Since the search for \{1 ... MaxD \} means the minimum and maximum number of DBSs, we first calculate the maximum number of DBS ($Max_D$), and then run the proposed algorithm in this interval.
\subsection*{Predict users}
\begin{figure}[h]
\begin{subfigure}{0.95\columnwidth}
\includegraphics[width=0.45\columnwidth,center]{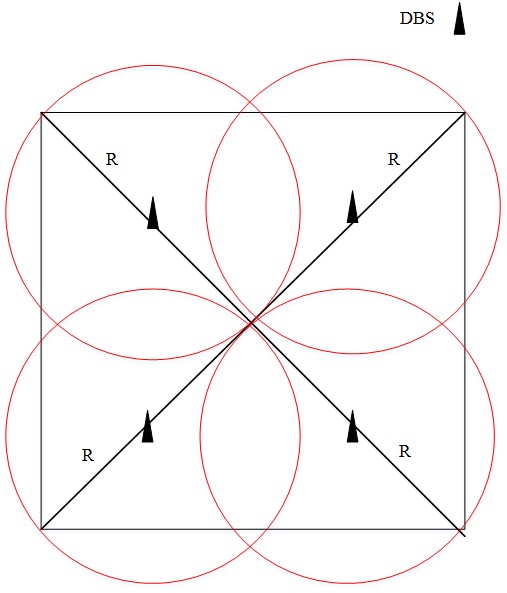}
\caption{with assumption of locating DBSs in these places, the total area is covered}
\label{fig.4a}
\end{subfigure}
\begin{subfigure}{0.95\columnwidth}
\includegraphics[width=0.45\columnwidth,center]{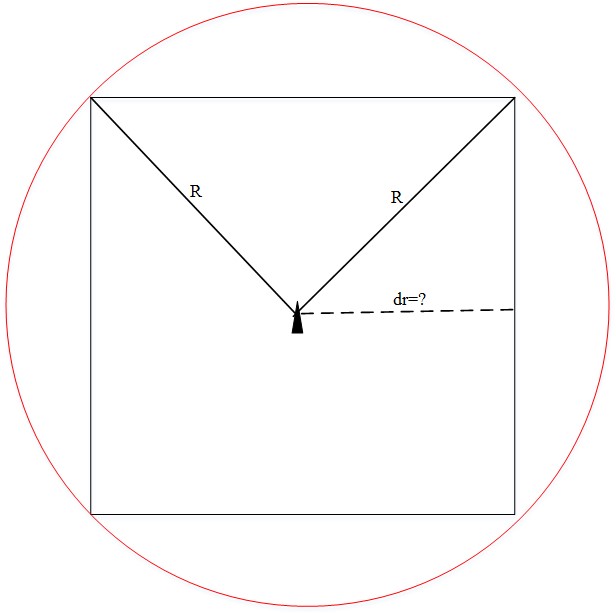}
\caption{the DBS coverage range}
\label{fig.4b}
\end{subfigure}
\caption{the DBS coverage model and coverage area }
\label{fig.4}
\end{figure}
As it can be seen from Figure \ref{fig.4a}, if DBSs which have $R$ coverage radius are located at the points indicated in the figure, the entire area with X and Y dimensions ($m$) will be fully covered. The maximum number of DBSs ($Max_D$) is calculated as follows:
\begin{equation} \label{eq.9}
\begin{split}
 & (2*dr)^2 = R^2 + R^2 \\
 & 4*dr^2=2R^2 \\
 & dr=\dfrac{R}{\sqrt{2}}
\end{split}
\end{equation}

\begin{equation} \label{eq.10}
\begin{split}
& Min_D = 1 \\
 & Max_D = \dfrac{X}{\dfrac{2*R}{\sqrt{2}}} +\dfrac{Y}{\dfrac{2*R}{\sqrt{2}}}
\end{split}
\end{equation}
After calculating the $Max_D$, the position of a target value within $(1,Max_D)$ is found and then the middle element $\beta_{\backepsilon}$  of the $(1, Max_D)$ is calculated as follows:
\begin{equation}\label{eq.11}
\mathop{\beta_{\backepsilon} = \dfrac{Min_D + Max_D}{2}}
\end{equation}
Following by that, we try to solve the problem statemen \eqref{eq.5} by using the $\beta_{\backepsilon}$ value to know whether the target area is covered effectively with $\beta_{\backepsilon}$ number of DBSs or not. If the all network users are covered with $\beta_{\backepsilon}$  number of DBSs (the problem is solved), then it is quite possible that full coverage is provided using less number of DBSs than $\beta_{\backepsilon}$. As a result, the search continues on the first half $(1…\beta_{\backepsilon})$ until it is successful. Whereas on the other hand, if using $\beta_{\backepsilon}$ number of DBSs, all the users are not covered, it is more likely to achieve more number of DBSs which are fully covered. Then, the search will be continued on the second half $(\beta_{\backepsilon}+1 ...  Max_D)$, until the problem is solved. 
Finally, if $\beta^{*}$ is less than $\alpha=0.9$, the number of DBSs will be increased; otherwise, the number of DBSs will be reduced. The proposed algorithm with a maximum number of iterations, $M_{\mu}$ can be summarized in algorithm \ref{alg1}.

\begin{algorithm}
\caption{An algorithm with caption}\label{alg1}
\begin{algorithmic}[1]
\Ensure $\beta_1 = 1 , \beta_2 = Max_D$
\State $M \gets 0$
\While{$M \leq M_{\mu}$} \Comment{$M_{\mu}:$ Maximum repetition of algorithm}
    \State $ \beta_{\backepsilon} \gets \dfrac{\beta_1 + \beta_2}{2}$
\If{problem is solved}
    \State $\beta^* = \beta^2 $
    \State \textbf{Break}
\EndIf
\State $M \gets M+1$
\If{Users are covered by $\beta_{\backepsilon}$ DBSs }
    \State $\beta_2 = \beta_{\backepsilon}$ \Comment{The search continues on the first half $(1,\beta_{\backepsilon})$}
\Else
    \State $\beta_1 = \beta_{\backepsilon}$\Comment{The search continues on the second  half $(\beta_{\backepsilon}+1,Max_D)$}
\EndIf
\If{$\beta^* < \alpha $ }
    \State $\beta_{\backepsilon} = \beta_{\backepsilon} + 1$  \Comment{increase the number of DBSs}
\Else
    \State $\beta_{\backepsilon} = \beta_{\backepsilon} - 1$  \Comment{decrease the number of DBSs}
\EndIf
\EndWhile
\end{algorithmic}
\end{algorithm}

\begin{figure}[H] 
    \centering
    \includegraphics[width =0.95\columnwidth] {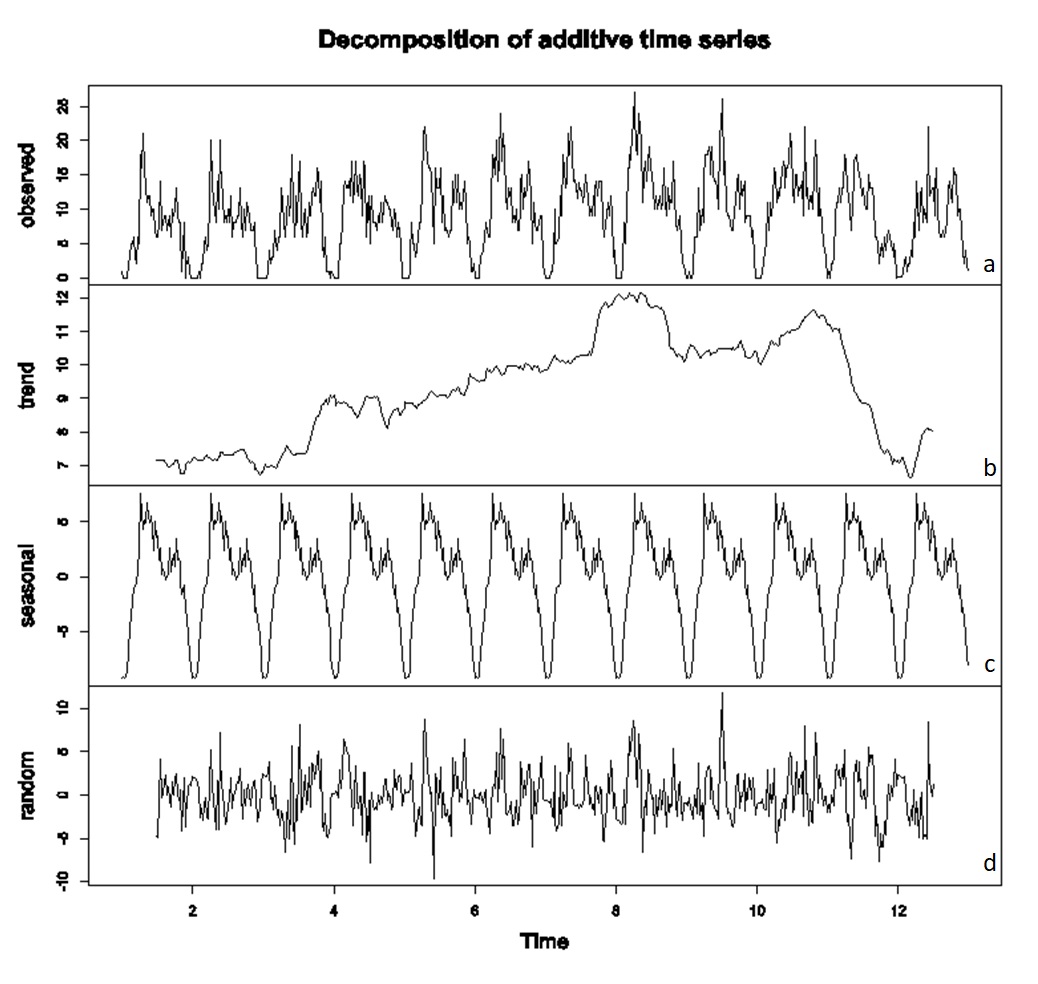}
    \caption{a) Main data, b) Cycle process, c) Seasonal component and d) Remaining}
    \label{fig.5}
\end{figure}

\subsection{Predict users}\label{sec.4.2}
\begin{figure}[H] 
    \centering
    \includegraphics[width =0.95\columnwidth] {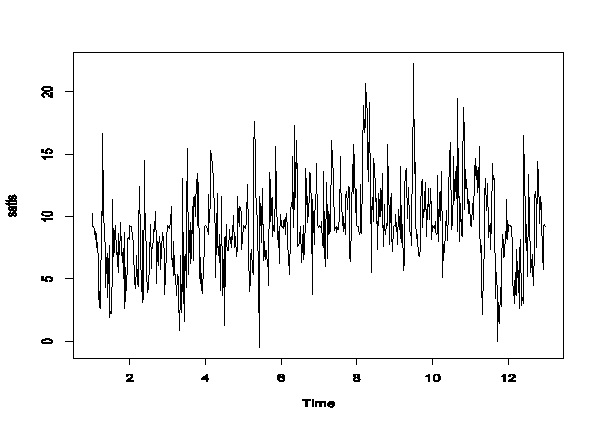}
    \caption{The remaining chart}
    \label{fig.6}
\end{figure}

Currently, we have the BS number at the moment t with the number of active users. In fact, we have used the real data of requests from users of the university's Internet network, which includes the spatial coordinates of each BS, the mac address, the number of online users covered by each The BS at different times includes the moon, day, hour, and minute. By predicting the number of users, each BS is predicted at the $t + 1$ moment.
Many of the prediction methods are based on the concept that, when there is a pattern in a time series data, that pattern can be determined by smoothing (averaging) the values of a random pattern. The effect of this smoothing is to eliminate randomness so that this pattern can be a model for the future and is used in prediction. In many cases, this pattern can be decomposed into sub-patterns that identify the time series components separately. Such an analysis can often be effective in understanding better the behavior of the series and increasing the accuracy of prediction.
Decomposition methods usually seek to identify distinct two-folds in the base model. These components are called seasonal factors and cyclic trends. Seasonal factor is associated with alternating fluctuations whose length is constant due to factors such as temperature, precipitation, moon, vacation time, and corporate policy. The trend-cyclical factor represents long-term changes at the level of the series.
In the principles of decomposition, it is assumed that the data are constructed as follows:
$Error + Template = Data$
$f = (Error, seasonal, trend-cycle) $
Therefore, in addition to the components of the pattern, it is assumed that there is an error or random element. In fact, this error is the difference between the effect of the combination of the two sub-patterns and the actual data. Therefore, it is often called an irregular component or the remainder. Figure \ref{fig.5} (a) shows the main data graph (the number of online users on Saturdays), (b) (trend-cycle diagram, (c) the seasonal component diagram, and Figure \ref{fig.6}remains the chart.

\subsection*{\textit{\textbf{Seasonal Moderation}}} 
A parsing consequence is to provide an easy way to calculate seasonally adjusted data. For a collective breakdown, seasonally adjusted data are simply calculated by subtracting the seasonal component.
\begin{equation}\label{eq.12}
\mathop{Y_t - S_t = T_t + E_t}
\end{equation}

\subsection*{\textit{\textbf{Prediction}}} 
The classification of Predicting methods is specified in two groups. A group called mediocre methods is consistent with the conventional mean concept, that is, observations having equal weights. The second group of methods allocates a set of unequal weights to past data, and because the weights in an exponential matrix are tangled from the newest to the furthest data, they are known as smoothing methods. All methods in the second group need to define the parameters to be private, and the values of these parameters lie between zero and the other. (These parameters determine the unequal weights that are assigned to the past data.) The hint-winters method consists of three smoothing parameters to smooth the data, trend, and seasonal indicator.
\subsection*{\textit{\textbf{Halt-Winters Prediction}}} 
The method of the hole was presented by Winters (1960) in order to directly consider the seasonal component. The hint-winters method is based on three equalization equations, one for the surface, one for the process, and one for the seasonal component.
This method is similar to the whole method, with an additional equation to deal with the seasonal component. Indeed, depending on whether the seasonal component is modeled collectively or in a multiplicative manner, there is a different duality of the \emph{Htl-Winters}.
In this application we have used a collective seasonal component that is as follows:
The main equations for the Holt-Winters collective method are:
\begin{equation}\label{eq.13}
\mathop{Level: L_t = \alpha (Y_t-S_{t-s}) + (1-\alpha)(L_{t-1} + b_{t-1}) }
\end{equation}
\begin{equation}\label{eq.14}
\mathop{Period: b_t = \beta (L_t-L_{t-1}) + (1-\beta)b_{t-1} }
\end{equation}
\begin{equation}\label{eq.15}
\mathop{Seasonal: S_t = \Upsilon (Y_t-L_t) + (1-\Upsilon)S_{t-s}}
\end{equation}
\begin{equation}\label{eq.16}
\mathop{Prediction: F_{t+m} = L_t + b_t m+S_{t-s+m} }
\end{equation}
In which, $s$ is the length of the season; $L_t$ indicates series level, $b_t$ shows the trend, $S_t$ for seasonal components,$F_{t+m}$ for prediction and $m$ for next step. As with all exponential smoothing methods, we need the initial values of the components to start the algorithm. In order to initialize the halt-winters prediction method, the initial values of $L_t$ the surface, $b_t$ trend, and $S_t$ seasonal indicator are required. The amount of the first level is obtained by averaging the first season:
\begin{equation}\label{eq.17}
\mathop{L_s = \dfrac{1}{s}(Y_1+Y_2+Y_3+...+Y_s)}
\end{equation}
Note that this moving average is s, so it eliminates the seasonality of the data. For initialization to the process, the use of two complete chapters (ie, two periods of s) is suitable as follows:
\begin{equation}\label{eq.18}
\mathop{b_s = \dfrac{1}{s}(\dfrac{Y_{s+1}-Y_1}{s}+\dfrac{Y_{s+2}-Y_2}{s}+...+\dfrac{Y_{s+s}-Y_s}{s})}
\end{equation}
Finally, for the initial values of seasonal indicators, we use the following expressions:
\begin{equation} \label{eq.19}
\begin{split}
 & S_2 = Y_2-L_s \\
 &\qquad   . \\
 &\qquad   . \\
 &\qquad   . \\
 & S_s = Y_s-L_s
\end{split}
\end{equation}

\begin{figure}[H] 
    \centering
    \includegraphics[width =0.95\columnwidth] {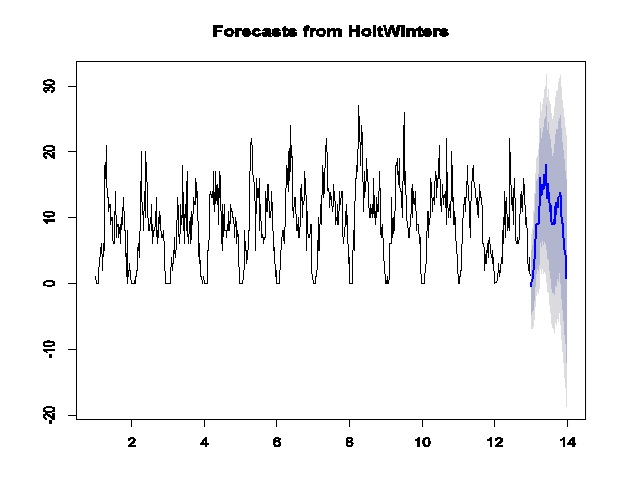}
    \caption{The number of online users predicted on Saturday afternoon}
    \label{fig.7}
\end{figure}

Now, we anticipate the number of online users on Saturday using this method and software $R$. In figure \ref{fig.7}, the central line represents a point prediction and a lower area represents a distance prediction with a 95\% confidence interval and a tall area representing a distance prediction with a confidence coefficient of 80\%.

\begin{figure}[H] 
    \centering
    \includegraphics[width =0.95\columnwidth] {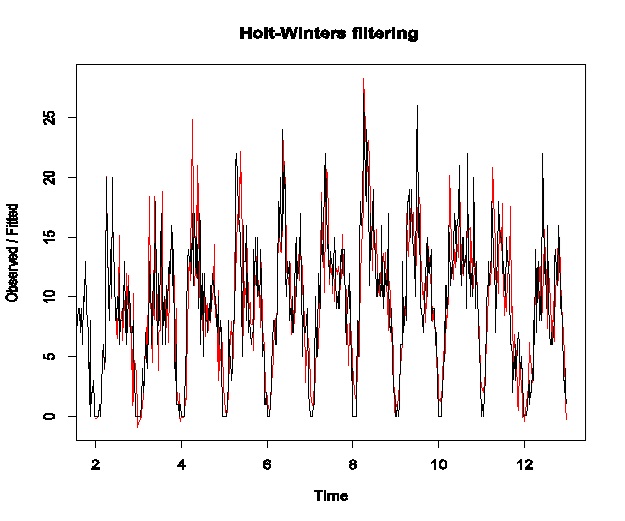}
    \caption{The actual data difference with the number of online users predicted with the help of the Hint Winters method}
    \label{fig.8}
\end{figure}

To check the accuracy of prediction done using the Hint Winters method, we can call this model on the same interval as available serial data, and see its mismatch with the main data in figure \ref{fig.8}.
 
In figure \ref{fig.8}, the red hint of the Winters model is a graph that, as it is known, is very close to the log data, and this prediction is acceptable. Also, for further examination, we anticipate the accuracy of the prediction of 12th Saturday based on 11th Saturday, which is as follows: \\*
{[1]} -0.7396121  0.1573054  0.5123411  1.3770198  1.9974246  3.5387841 \\*
{[7]}  7.3981624  8.0500743  9.2047099  9.8508378  9.2753863  9.4302746 \\*
{[13]}  9.9756647 12.6505924 15.3198268 14.3176929 14.1138053 14.4615613 \\*
{[19]} 14.1592316 15.4278365 17.4821584 15.8422638 15.1756764 15.0669139 \\*
{[25]} 14.2271335 13.4751354 15.9438215 14.2027739 13.3746902  9.9046706 \\
{[31]} 10.9922352  9.9410336 10.3844131  9.7595742  9.5116178  9.6961117 \\
{[37]}  8.8208381 12.1951457  9.9951535 11.1819512 11.1907276 11.5572057 \\
{[43]} 13.6670028 12.2485367 14.0621561 13.7516646 11.3720449  9.1870023 \\
{[49]}  9.3968195  7.5904347  7.2562910  6.1377897  4.1359894  2.4946793 \\
{[55]}  0.5017181 \\

\begin{figure}[H] 
    \centering
    \includegraphics[width =0.95\columnwidth] {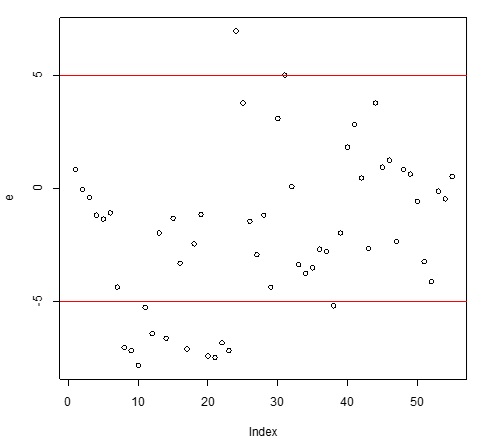}
    \caption{The difference between the actual data and the number of online users predicted using the Hint Winters method}
    \label{fig.9}
\end{figure}

And we have the actual data of the twelfth Saturday, which is as follows: \\
{[1]}  0.09  0.09  0.09  0.18  0.64  2.45  3.00  1.00  2.00  2.00  4.00  3.00 \\
{[13]}  8.00  6.00 14.00 11.00  7.00 12.00 13.00  8.00 10.00  9.00  8.00 22.00 \\ 
{[25]} 18.00 12.00 13.00 13.00  9.00 13.00 16.00 10.00  7.00  6.00  6.00  7.00 \\
{[37]}  6.00  7.00  8.00 13.00 14.00 12.00 11.00 16.00 15.00 15.00  9.00 10.00 \\
{[49]} 10.00  7.00  4.00  2.00  4.00  2.00  1.00 \\
And by the difference of the two sets, the model error value for the estimation of the twelfth of Saturday is based on the Healt Winters method on the previous eleventh of the previous day: \\
{[1]}  0.82961215 -0.06730540 -0.42234112 -1.19701976 -1.35742461 -1.08878411 \\
{[7]} -4.39816239 -7.05007431 -7.20470986 -7.85083781 -5.27538634 -6.43027462 \\
{[13]} -1.97566470 -6.65059237 -1.31982676 -3.31769286 -7.11380534 -2.46156131 \\
{[19]} -1.15923161 -7.42783651 -7.48215841 -6.84226378 -7.17567638  6.93308607 \\
{[25]}  3.77286655 -1.47513537 -2.94382146 -1.20277386 -4.37469019  3.09532943 \\
{[31]}  5.00776483  0.05896641 -3.38441313 -3.75957423 -3.51161778 -2.69611168 \\
{[37]} -2.82083807 -5.19514570 -1.99515351  1.81804880  2.80927242  0.44279435 \\
{[43]} -2.66700283  3.75146326  0.93784394  1.24833538 -2.37204493  0.81299770 \\
{[49]}  0.60318049 -0.59043470 -3.25629099 -4.13778971 -0.13598944 -0.49467925 \\
{[55]}  0.49828185
By drawing an error diagram (as shown in Figure \ref{fig.9}), we find that the errors are scattered around zero and do not follow any particular process.
 After calculating the number of users in the next moment, we calculate the optimal number and location of the DBS. The optimal DBS location and location in the current state and the next, with the assumption of minimizing the pathloss, are calculated in Sections \ref{sec.4.1} and \ref{sec.4.2}, respectively. Now we will optimize the DBS transfer from current to next mode in section \ref{sec.4.3}.
\subsection{Optimizing Transfer (Mobility DBS)}\label{sec.4.3}
\begin{figure}[h]
\begin{subfigure}{0.5\textwidth}
\includegraphics[width=0.45\linewidth,center]{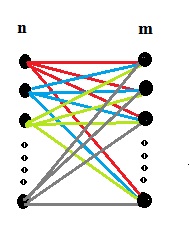}
\label{fig.10a}
\end{subfigure}
\begin{subfigure}{0.5\textwidth}
\includegraphics[width=0.45\linewidth,center]{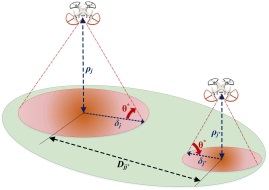}
\label{fig.10b}
\end{subfigure}
\caption{Possible scenarios of DBS transfer from current location to predicted locations}
\label{fig.10}
\end{figure}

Suppose $n$ has calculated the candidate location for the presence of DBS (in Section \ref{sec.4.1}) at the present moment, so that
\begin{equation}\label{eq.20}
\mathop{n \in \{1,...NDBS_n\} DBS[n]}
\end{equation}
We have obtained the candidate's place for the presence of DBS at the next moment (according to Section \ref{sec.4.2}), so that
\begin{equation}\label{eq.21}
\mathop{m \in \{1,...NDBS_n\} DBS[m]}
\end{equation}
We must find the best transmission between the elements of the two sets. Possible transfers between these two modes are shown in figure \ref{fig.10}.

We analyze the transfer problem optimally as follows:
\begin{equation}\label{eq.22}
\mathop{ \sum_{i=1}^{n}\sum_{j=1}^{m} d_{i,j} * dis_{i,j} }
\end{equation}
Subject to:

\begin{equation}\label{eq.23}
\mathop{ \sum_{i=1}^{n} d_{i,j}=1 }
\end{equation}

\begin{equation}\label{eq.24}
\mathop{ \sum_{j=1}^{m} d_{i,j}=1 }
\end{equation}

\begin{equation}\label{eq.25}
  d_{i,j} =
    \begin{cases}
      1 & \text{if the transition between $i$ adn $j$ is selected}\\
      0 & \text{otherwise}
    \end{cases}       
\end{equation}

First, to optimize the transfer, we calculate the distance between each point of the two sets $dist_{i,j}$ based on the speed of the DBS movement and the time.In fact, we check that if, given the speed of DBS and the transfer time, the DBS can move from its place $i$ to $j$, then the distance between the two locations is calculated according to Euclidean, otherwise we set the $dis_{i,j}$ value to $\infty$. \\
Hamelen, as stated in equation \eqref{eq.23}, $dist_{i,j}$ is the distance between the place $i$ at the instant moment and the place $j$ at the next moment, and the transition selection index is that if $i$ is transferred to $j$, its value 1 otherwise is 0 (according to equation \eqref{eq.25}). Equation \eqref{eq.23} and \eqref{eq.24} examine one-to-one relationship, so that from $n$ to $m$, we have only one transition. Finally, if $n = m$, then the problem is solved.
\begin{equation} \label{eq.26}
\begin{split}
 & \forall j \sum_{i=1}^{n}d_{i,j}\leq 1 ,\quad if n<m\\
 & \forall i \sum_{j=1}^{m}d_{i,j}\leq 1 ,\quad if n>m
\end{split}
\end{equation}
To solve this problem, we use a Simplex method \cite{bertsimas-LPbook}.	 

\section{Evaluation}\label{sec.5}
In this section, we evaluate the method of DBS placement and the minimum number of them based on a real university data. First, we introduce the parameters used in simulation in Section \ref{sec.5.1}. Then we will provide details about the actual data used in the simulation in Section \ref{sec.5.2}. Subsequently, the evaluation criteria are described in Section \ref{sec.5.3} and then the output results are presented in the diagram format in Section \ref{sec.5.4}.

\subsection{Simulation Setup}\label{sec.5.1}
Consider a university campus area of 300 * 400 meters which the number of subscribers of each BS will be considered as the input. A number of users are randomly distributed in the area (in different scenarios, different user numbers are used). The altitude of the DBSs and their coverage radius are constant values, 50 and 100 meters, respectively. By taking advantage of OPL software, we try to simulate and optimize the problem. Simulation parameters are listed in Table \ref{tab1}.

\begin{table}[h]
\begin{center}
    \caption{Simulation Parameters}
    \label{tab1}   
    
\begin{tabularx}{0.95\columnwidth} { 
  | >{\centering\arraybackslash}X 
  | >{\centering\arraybackslash}X 
  | >{\centering\arraybackslash}X | }
\hline
 \textbf{Variable} & \textbf{description} & \textbf{Values} \\
 \hline
 $R$ & DBS coverage range & 50 M \\
  \hline
 $H$ & Height & 100M \\
  \hline
 $X, Y$ & Width and length of target area & 300* 400 \\
  \hline
 $NU$ & Number of users & Various numbers \\
  \hline
 $U_{bw}$ & Maximum required bandwidth of users & Is randomly selected from \{0.1,0.5,1,1.5,2\} \\
  \hline
 $Max_D$ & Maximum Bandwidth of DBSs & (10,20,50,100) \\
 \hline
\end{tabularx}
\end{center}
\end{table}

\subsection{Simulation results}\label{sec.5.2}
The following diagrams are discussed in this article:

\begin{itemize}
    \item Percentage of users covered by DBS at the moment $t$ and $t+1$.
    \item Optimal transfer of DBS from moment and current location to next (transfer from moment $t$ to $t+1$. 
    \item Sum-Rate DBS service for users (bandwidth users) at the moment $t$ and $t+1$.
    \item Spatial distribution and the appropriate number of DBS to cover the variable number of users.
\end{itemize}

\subsection{Evaluation Criteria }\label{sec.5.3}

We have achieved results for 4 categories of users (50, 100, 150, 200) and 4 DBS types with a maximum bandwidth of \{10,20,30,40\} Mb. In order not to prolong the article, only the results related to 150 users will be further reviewed.
\begin{figure}[H] 
    \centering
    \includegraphics[width =0.95\columnwidth] {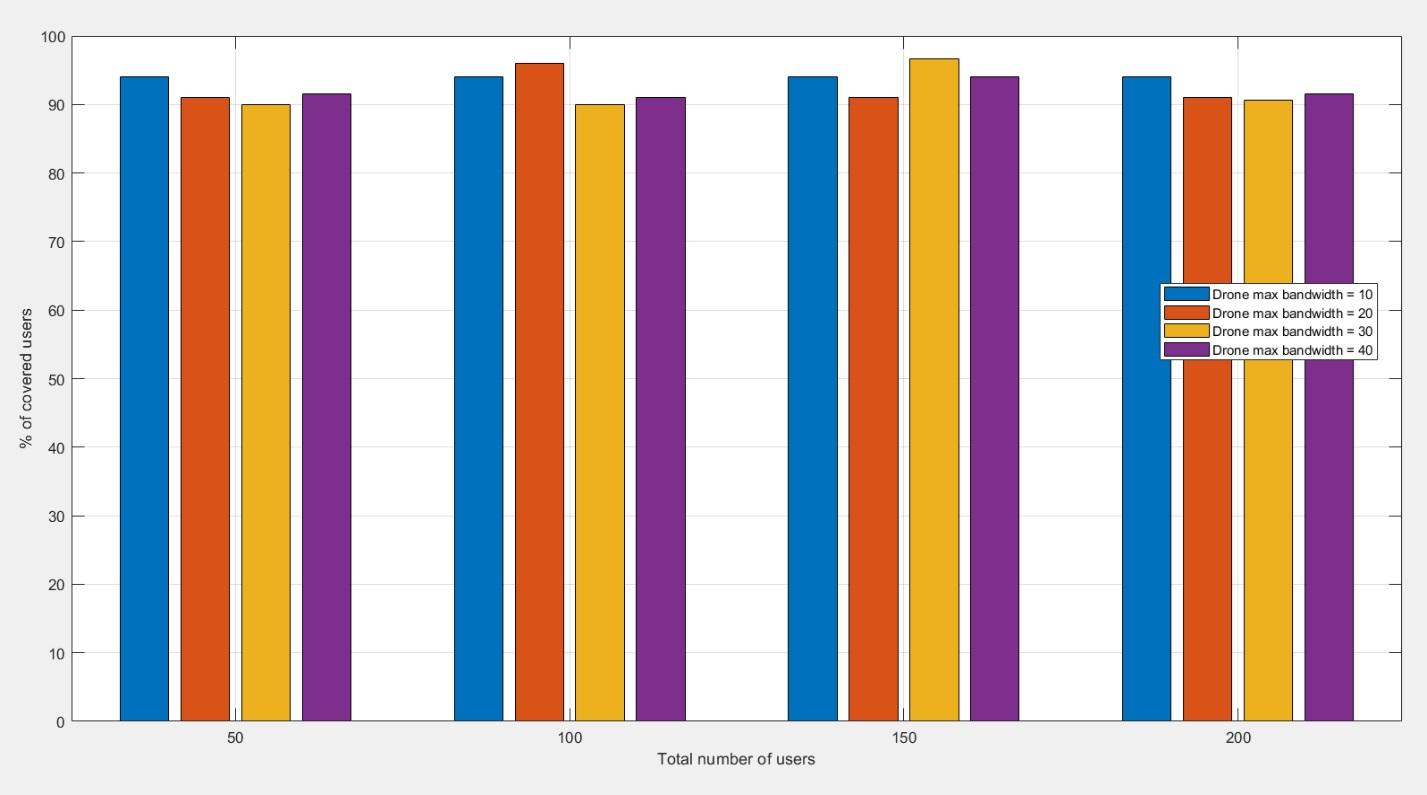}
    \caption{Percentage of users covered by 4 types of DBS with maximum bandwidth (10,20,30,40) Mbps}
    \label{fig.11}
\end{figure}

In Figure \ref{fig.11}, the percentage of users covered by DBS with different bandwidths is shown. We run the simulation several times with different DBS with bandwidths (10,20,30,40) Mbps. As shown in Figure \ref{fig.11}, the number of users and the maximum bandwidth of DBS affect the final result (coverage percentage). Figure \ref{fig.11} uses DBS with a maximum bandwidth of 40 MB, 93\% of the 50 users, 91\% of 100 users, 95\% of the 150 users and about 92\% of the 200 users. Similarly, if using DBS with a maximum bandwidth of 10 megabytes to cover users, about 95\% of the 50, 100, 150, and 200 users will be served by this DBS.
\begin{figure}[H] 
    \centering
    \includegraphics[width =0.95\columnwidth] {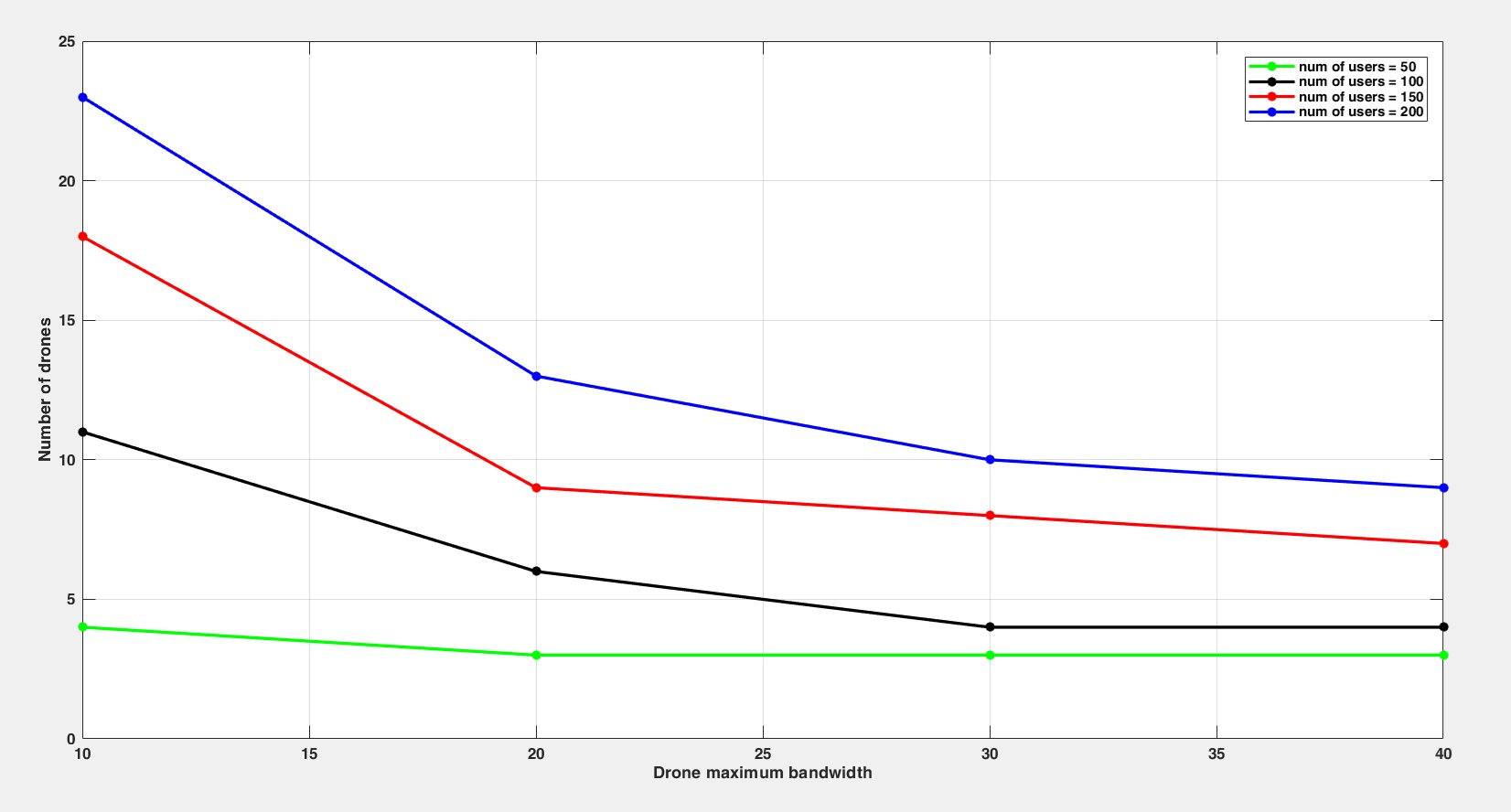}
    \caption{There are 4 different DBS types with a maximum bandwidth assuming a fixed number of users (50, 100, 150, and 100)}
    \label{fig.12}
\end{figure}
We run the simulation several times on a number of different users to find the required number of required DBS with the maximum bandwidth of the variable (10 to 40 MB). As you can see in Figure \ref{fig.12}, in general, the smaller the number of users, the less DBS is needed to cover and service. On the other hand, there is an inverse relationship between the maximum bandwidth and the required number of DBSs, so that with the increase in the maximum bandwidth of DBS in each set of users, a lower number of DBSs is required. As shown in Figure \ref{fig.12}, for the coverage of 50 users to 4 and for 200 users, six times the number, 24 DBS with a maximum bandwidth of 10 Mb, on the other hand for the same number of users, respectively, to 3 and 9 DBS with the maximum bandwidth 40 MB needed.
\begin{figure}[H] 
    \centering
    \includegraphics[width =0.95\columnwidth] {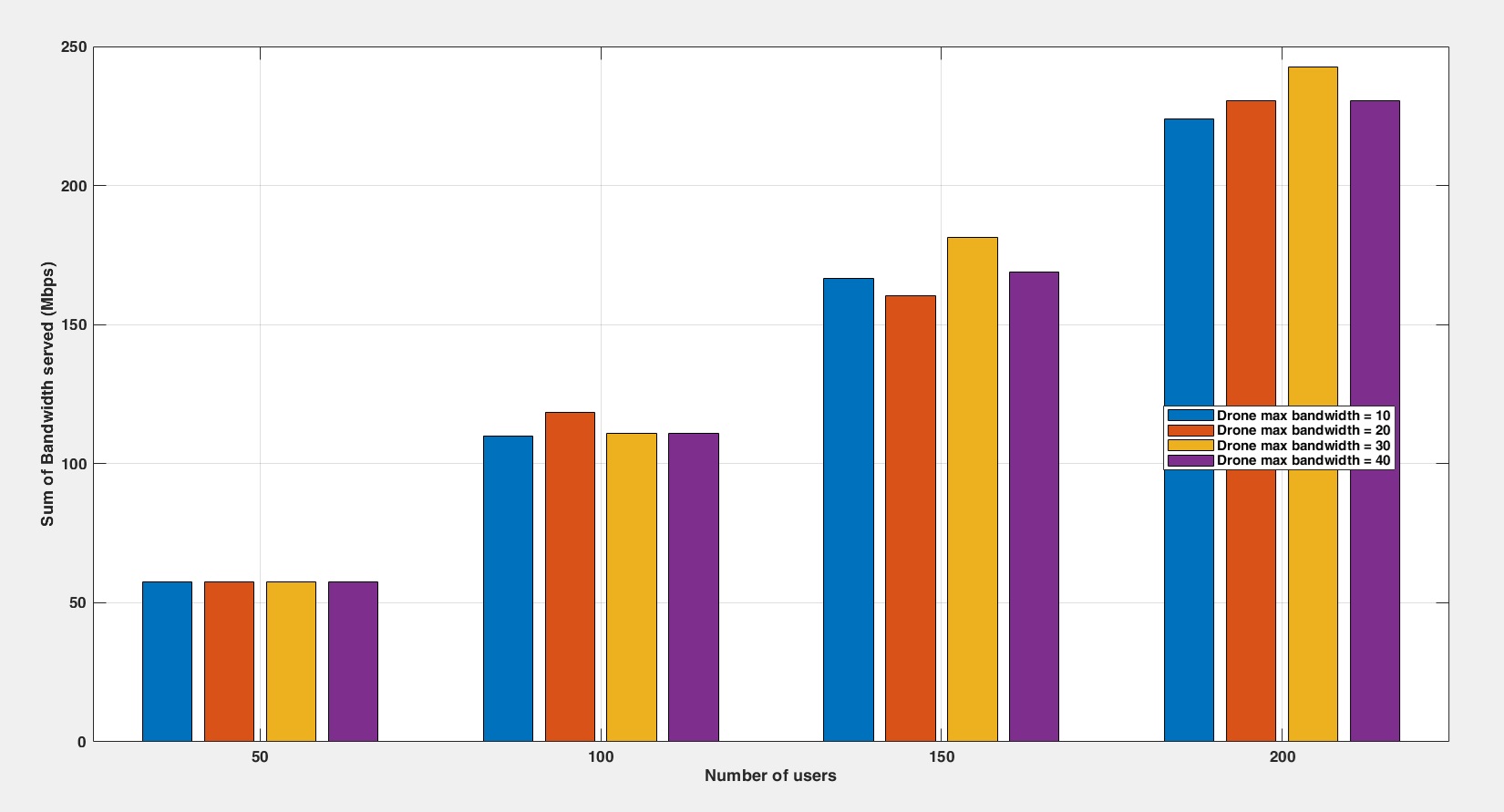}
    \caption{Sum-Rate service based on the maximum bandwidth of various DBS (10,20,30,40) Mbit/s}
    \label{fig.13}
\end{figure}
Figure \ref{fig.13} shows the sum-rate service for different users for the number of DBS variables with the maximum bandwidth (10,20,30,40) Mbit/s. As seen from the figure, the Sum-Rate service is directly related to the maximum bandwidth of the DBS to cover any user-defined number. In other words, the Sum-Rate chart, which offers DBS with a maximum bandwidth of 40 Mbps, is more than Sum-Rate's DBS service with a maximum bandwidth of 10 Mbps. According to the figure, the highest Sum-Rate service is provided at around 240 Mbps, which is provided by DBS with a maximum bandwidth of 30 Mb to cover 200 users and the lowest Sum-Rate service provided is approximately 60 Mbps for serving 50 users by DBS With a maximum bandwidth of 10 megabits.
\begin{figure}[H] 
    \centering
    \includegraphics[width =0.95\columnwidth] {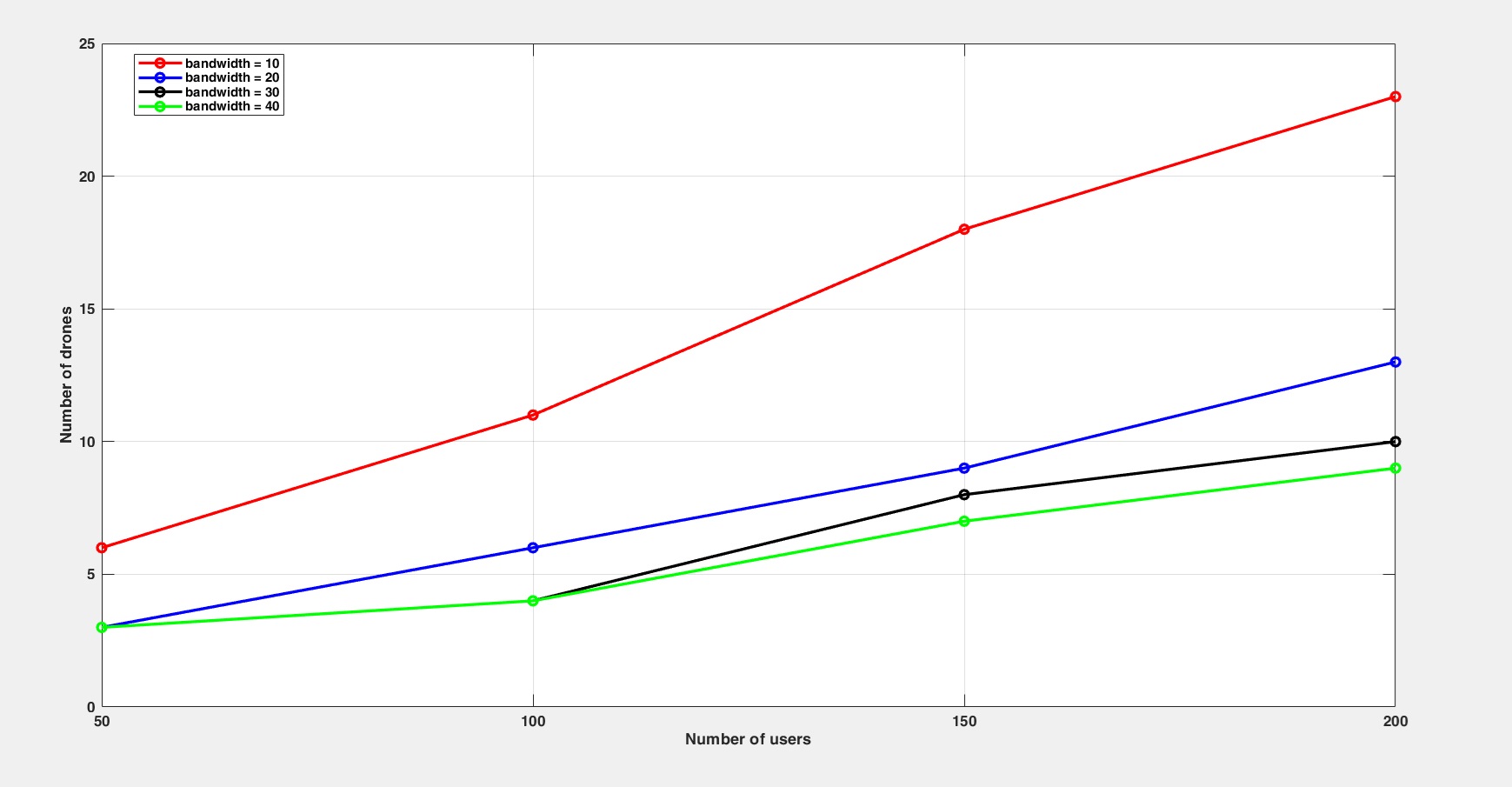}
    \caption{Number of DBSs per user variable}
    \label{fig.14}
\end{figure}
Figure \ref{fig.14} shows the required number of DBS with maximum bandwidths of {10,20,30, 40} Mbit in order to temporarily cover the number of users in the range of 50 to 200. It does not change the accuracy of the location of the previous users, and each time a number of new users are added to that environment. As shown in figure \ref{fig.14}. By increasing the number of users, more DBSs (with each bandwidth) will be required to cover users. As per figure \ref{fig.14}, for 50 users 6 and for 200 users, it is required about 24 (four times) DBS with a maximum 10 Mbps bandwidth. Because as the number of users increases, users may be more dispersed in the environment, hence the need for more DBS coverage is required. Additionally, in proportion to the maximum DBS bandwidth, more users are served by that DBS. For example, as shown in figure \ref{fig.14}, to cover 150 users, 18 DBSs with a maximum bandwidth of 10 megabytes or 9 (half) DBSs with a maximum bandwidth of 40 MB are required. A summary of the information in figure \ref{fig.14} is shown in table \ref{tab2}.

\begin{table}[h]
\centering
    \caption{Required DBS with bandwidth (10,20,30,40) for number coverage (50,100,150 and 200) users.}
    \label{tab2}   
    
\begin{tabularx}{0.95\columnwidth} { 
  | >{\centering\arraybackslash}X 
  | >{\centering\arraybackslash}X 
  | >{\centering\arraybackslash}X
  | >{\centering\arraybackslash}X 
  | >{\centering\arraybackslash}X | }
\hline

Number of Users & \multicolumn{4}{|c|}{Number of DBSs with }\\\cline{2-5}
 & 100 MB & 50 MB  & 20 MB & 10 MB \\
\hline
50  & 5  &  5 & 7  & 8  \\
\hline
100 & 5  &  5 & 9  & 14  \\
\hline
150 & 9  & 10 & 12 & 18  \\
\hline
200 & 13 & 13 & 16 & 24 \\
\hline
\end{tabularx}
\end{table}

\subsection{Results Diagram}\label{sec.5.4}

\begin{figure}[ht] 
  \begin{subfigure}[b]{0.5\linewidth}
    \centering
     \captionsetup{margin=0.2cm}
    \includegraphics[width=0.75\linewidth]{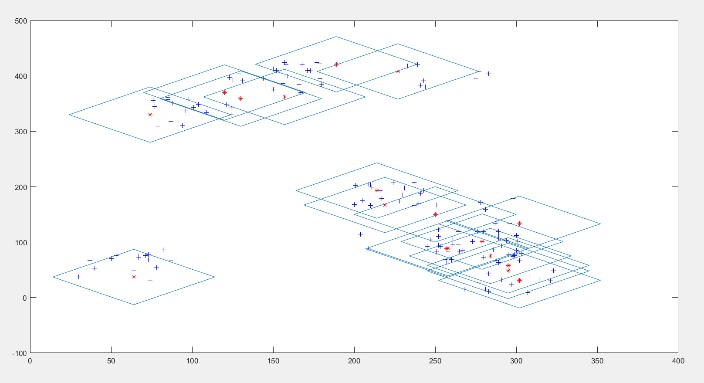} 
    \caption{Optimal location of 18 DBS with a maximum bandwidth of 10 Mbit} 
    \label{fig.15a} 
    \vspace{4ex}
  \end{subfigure}
  \begin{subfigure}[b]{0.5\linewidth}
    \centering
    \captionsetup{margin=0.2cm}
    \includegraphics[width=0.75\linewidth]{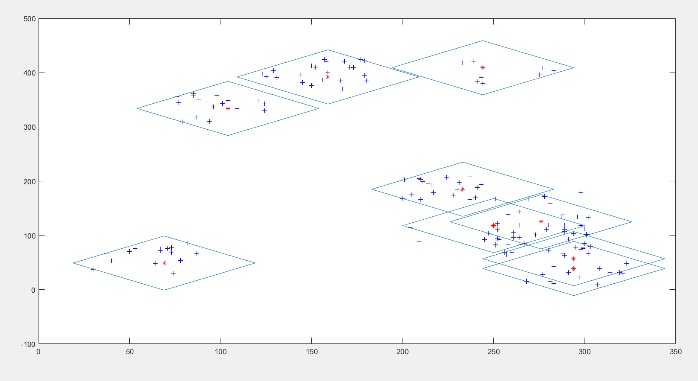} 
    \caption{Optimal location 9 DBS with a maximum bandwidth of 20 Mbit} 
    \label{fig.15b} 
    \vspace{4ex}
  \end{subfigure} 
  \begin{subfigure}[b]{0.5\linewidth}
    \centering
    \captionsetup{margin=0.2cm}
    \includegraphics[width=0.75\linewidth]{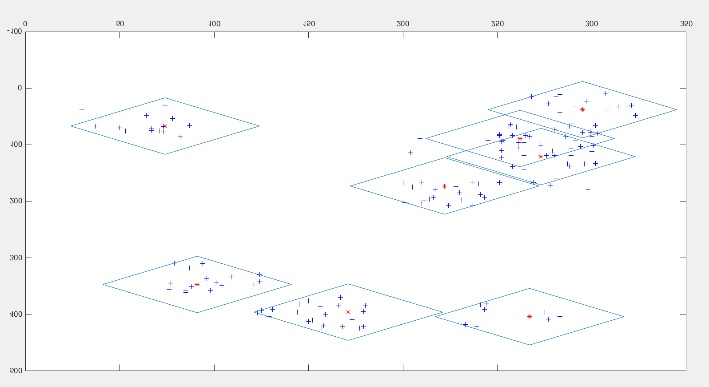} 
    \caption{Optimum location 8 DBS with maximum 30 Mbps bandwidth} 
    \label{fig.15c} 
  \end{subfigure}
  \begin{subfigure}[b]{0.5\linewidth}
    \centering
    \captionsetup{margin=0.2cm}
    \includegraphics[width=0.75\linewidth]{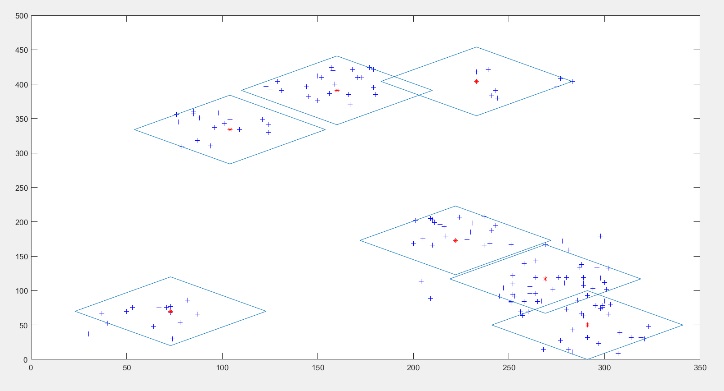} 
    \caption{Optimal location 7 DBS with 40 Mbps maximum bandwidth} 
    \label{fig.15d} 
  \end{subfigure} 
  \caption{Optimum location and minimum required DBS with maximum bandwidths (10,20,30,40) to cover 150 users at the instant t}
  \label{fig.15} 
\end{figure}

Figure \ref{fig.15} shows the location of 150 users and the BDSs with a maximum bandwidth of \{10,20,30 and 40\}Mb at the instant $t$. In this form, users with the $+$ (blue) sign, DBS with the $*$ (red) sign and the coverage area of each DBS are shown with blue rhombus, because in condition 6 of the pathloss, we use norm1 to calculate the distance. In figure \ref{fig.15a}, we have distributed 150 users in the environment, according to the proposed algorithm, the optimal number of DBSs with a maximum bandwidth of 10 Mb in order to maximize the coverage of these 150 users is 18, the best positioning of which is specified in the figure. Figure \ref{fig.15b} shows the scenario used to cover these 150 users with 9 DBS with 20 Mbps bandwidth, as shown in the figure where there are no user-defined DBSs. Figure \ref{fig.15c} shows the location of users and location of 8 DBSs with 30 Mbps bandwidth. In the last scenario, to cover this number of users, at least 7 DBSs with 40 Mbps bandwidth are required, their optimal location shown in figure \ref{fig.15d}. As seen from the figures, the number of dumps needed to cover these 150 users decreases with their maximum bandwidth increase.

\begin{figure}[ht] 
  \begin{subfigure}[b]{0.5\linewidth}
    \centering
     \captionsetup{margin=0.2cm}
    \includegraphics[width=0.75\linewidth]{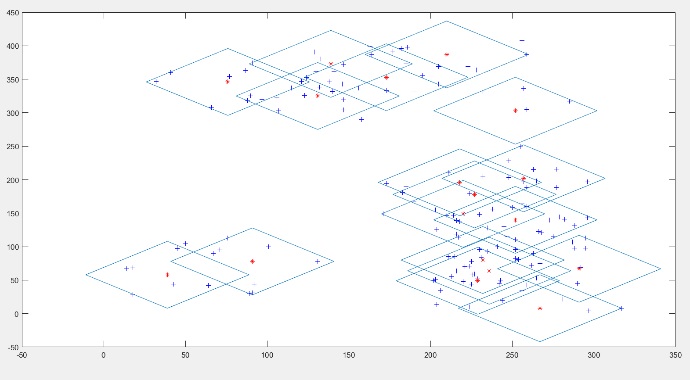} 
    \caption{Optimal location 18 DBS with a maximum bandwidth of 10 Mbit} 
    \label{fig.16a} 
    \vspace{4ex}
  \end{subfigure}
  \begin{subfigure}[b]{0.5\linewidth}
    \centering
    \captionsetup{margin=0.2cm}
    \includegraphics[width=0.75\linewidth]{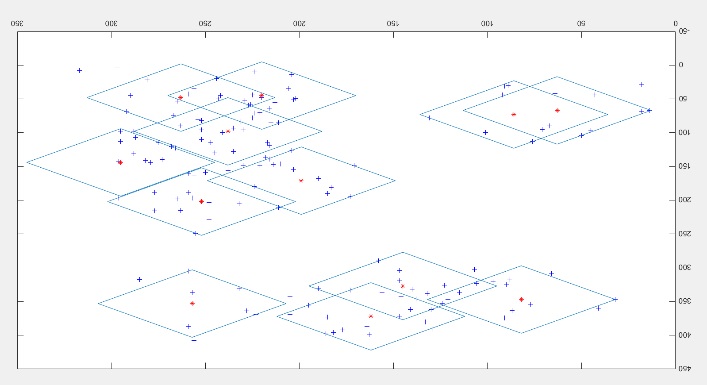} 
    \caption{Optimal location 12 DBS with a maximum bandwidth of 20 Mbit} 
    \label{fig.16b} 
    \vspace{4ex}
  \end{subfigure} 
  \begin{subfigure}[b]{0.5\linewidth}
    \centering
    \captionsetup{margin=0.2cm}
    \includegraphics[width=0.75\linewidth]{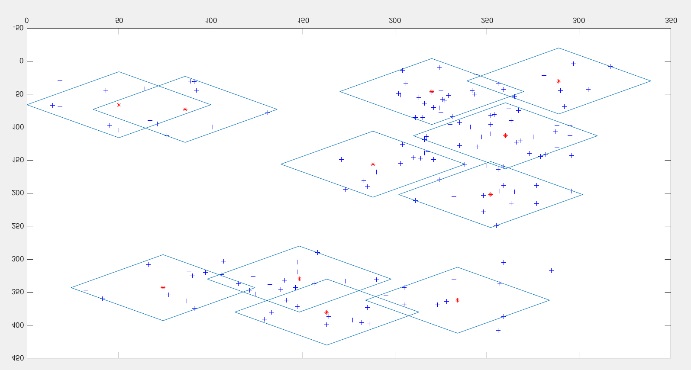} 
    \caption{Optimal location 11 DBS with a maximum bandwidth of 30 Mb} 
    \label{fig.16c} 
  \end{subfigure}
  \begin{subfigure}[b]{0.5\linewidth}
    \centering
    \captionsetup{margin=0.2cm}
    \includegraphics[width=0.75\linewidth]{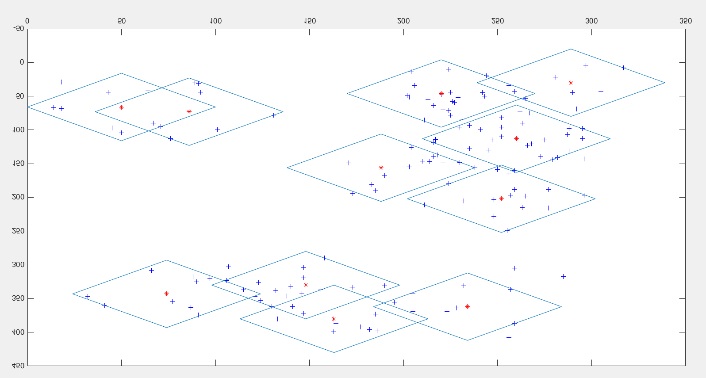} 
    \caption{Optimal location 11 DBS with a maximum bandwidth of 40 megabit} 
    \label{fig.16d} 
  \end{subfigure} 
  \caption{Optimal location and minimum number of required DBS with maximum bandwidths (10,20,30,40) to cover 150 users at the time t + 1}
  \label{fig.16} 
\end{figure}
Figure \ref{fig.16} shows the location and number of required BDS with the maximum bandwidth of \{10,20,30 and 40\}Mb in order to cover the same 150 users at the instant $t + 1$. In figure \ref{fig.16a}, for the coverage of 150 users At least 18 DBS with a maximum of 10 Mbps bandwidth are required, the optimal locations for this DBS are shown in the figure. As shown in figure \ref{fig.16b}, this number can be covered with 12 DBSs with 20 Mbps bandwidths. Figure \ref{fig.16c} and \ref{fig.16d}; respectively, the number and location of 11 DBSs with bandwidth of 30 and 40 Mbit shows the coverage of 150 users at the moment $t + 1$.

\begin{figure}[ht] 
  \begin{subfigure}[b]{0.5\linewidth}
    \centering
     \captionsetup{margin=0.2cm}
    \includegraphics[width=0.75\linewidth]{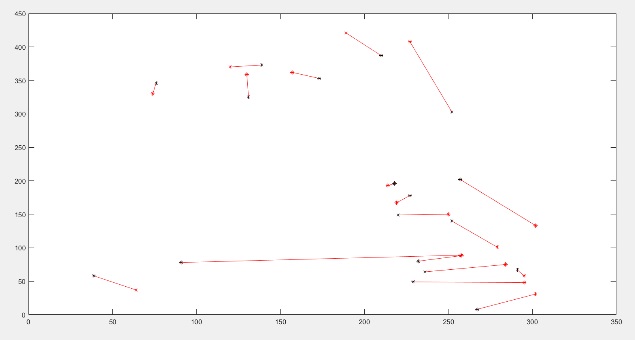} 
    \caption{Optimal transfer of DBS with a maximum bandwidth of 10 Mb from moment t to t + 1} 
    \label{fig.17a} 
    \vspace{4ex}
  \end{subfigure}
  \begin{subfigure}[b]{0.5\linewidth}
    \centering
    \captionsetup{margin=0.2cm}
    \includegraphics[width=0.75\linewidth]{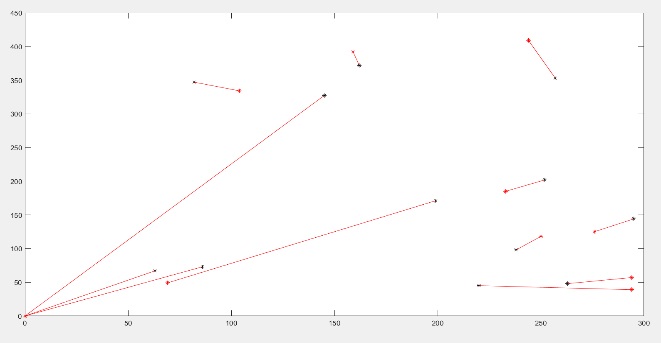} 
    \caption{Optimal DBS transmission with a maximum bandwidth of 20 Mb from moment t to t + 1} 
    \label{fig.17b} 
    \vspace{4ex}
  \end{subfigure} 
  \begin{subfigure}[b]{0.5\linewidth}
    \centering
    \captionsetup{margin=0.2cm}
    \includegraphics[width=0.75\linewidth]{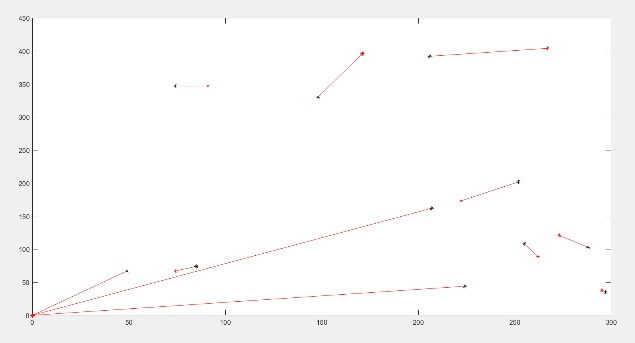} 
    \caption{Optimal transfer of DBS with a maximum bandwidth of 30 Mbit from moment t to t + 1} 
    \label{fig.17c} 
  \end{subfigure}
  \begin{subfigure}[b]{0.5\linewidth}
    \centering
    \captionsetup{margin=0.2cm}
    \includegraphics[width=0.75\linewidth]{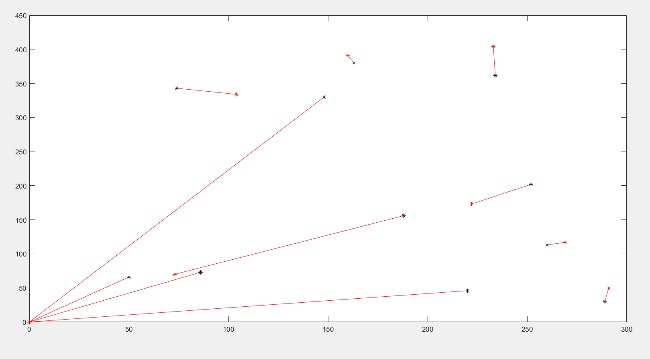} 
    \caption{ptimal transfer of DBS with a maximum bandwidth of 40 Mbit from moment t to t + 1} 
    \label{fig.17d} 
  \end{subfigure} 
  \caption{DBS transmission with maximum bandwidths of {10,20,30,40 mb} from moment t to t + 1.}
  \label{fig.17} 
\end{figure}

After selecting the optimal location and number at time t and predicting at $t + 1$, we choose the most optimal transfer from the possible transfers. Figure \ref{fig.17} shows optimal transfer of DBS from the current optimal location at the instant $t$ (shown with * red) to the optimal location at the moment $t + 1$ (* Black) with the goal of maximizing the number of covered users.
\section{Conclusion}\label{sec.6}
In the present paper, a new method is presented for optimally deploying and determining the minimum number of requirements for DBS in a region with different user density at the current and future moment, so that the maximum number of users is served and covered. Since computational complexity does not solve the problem of optimization with existing methods, we use a proposed algorithm to solve this nonlinear problem. In this method, the minimum number of DBS and their optimal location at the moment t and t + 1 are calculated and predicted based on the available information, assuming the path loss is minimized. Finally, the DBS transition is optimized from the current state to the predicted future locations. The simulation results show the acceptable performance of the proposed method despite the different density of users in the region. In other words, the number of DBS in the region is proportional to the density of users in that area.

\bibliographystyle{IEEEtran} 
\bibliography{Refrences}

\begin{thebibliography}{10}
\providecommand{\url}[1]{#1}
\csname url@samestyle\endcsname
\providecommand{\newblock}{\relax}
\providecommand{\bibinfo}[2]{#2}
\providecommand{\BIBentrySTDinterwordspacing}{\spaceskip=0pt\relax}
\providecommand{\BIBentryALTinterwordstretchfactor}{4}
\providecommand{\BIBentryALTinterwordspacing}{\spaceskip=\fontdimen2\font plus
\BIBentryALTinterwordstretchfactor\fontdimen3\font minus
  \fontdimen4\font\relax}
\providecommand{\BIBforeignlanguage}[2]{{%
\expandafter\ifx\csname l@#1\endcsname\relax
\typeout{** WARNING: IEEEtran.bst: No hyphenation pattern has been}%
\typeout{** loaded for the language `#1'. Using the pattern for}%
\typeout{** the default language instead.}%
\else
\language=\csname l@#1\endcsname
\fi
#2}}
\providecommand{\BIBdecl}{\relax}
\BIBdecl

\bibitem{Li2021}
L.~Li, X.~Wen, Z.~Lu, W.~Jing, and H.~Zhang, ``Energy-efficient multi-uavs
  deployment and movement for emergency response,'' \emph{IEEE Communications
  Letters}, vol.~25, pp. 1625--1629, 2021.

\bibitem{9586045}
Z.~Zhao, L.~Pacheco, H.~Santos, M.~Liu, A.~D. Maio, D.~Rosario, E.~Cerqueira,
  T.~Braun, and X.~Cao, ``Predictive uav base station deployment and service
  offloading with distributed edge learning,'' \emph{IEEE Transactions on
  Network and Service Management}, vol.~18, no.~4, pp. 3955--3972, 2021.

\bibitem{Mozaffari2019}
M.~Mozaffari, W.~Saad, M.~Bennis, Y.-H. Nam, and M.~Debbah, ``A tutorial on
  uavs for wireless networks: Applications, challenges, and open problems,''
  \emph{IEEE Communications Surveys Tutorials}, vol.~21, no.~3, pp. 2334--2360,
  2019.

\bibitem{8903295}
D.~Tezza and M.~Andujar, ``The state-of-the-art of human-drone interaction: A
  survey,'' \emph{IEEE Access}, vol.~7, pp. 167\,438--167\,454, 2019.

\bibitem{7962642}
E.~Kalantari, M.~Z. Shakir, H.~Yanikomeroglu, and A.~Yongacoglu,
  ``Backhaul-aware robust 3d drone placement in 5g+ wireless networks,'' in
  \emph{2017 IEEE International Conference on Communications Workshops (ICC
  Workshops)}, 2017, pp. 109--114.

\bibitem{Fotouhi2019}
A.~Fotouhi, H.~Qiang, M.~Ding, M.~Hassan, L.~G. Giordano, A.~Garcia-Rodriguez,
  and J.~Yuan, ``Survey on uav cellular communications: Practical aspects,
  standardization advancements, regulation, and security challenges,''
  \emph{IEEE Communications Surveys \& Tutorials}, vol.~21, pp. 3417--3442, 3
  2019.

\bibitem{Lai2019}
C.-C. Lai, C.-T. Chen, and L.-C. Wang, ``On-demand density-aware uav base
  station 3d placement for arbitrarily distributed users with guaranteed data
  rates,'' \emph{IEEE Wireless Communications Letters}, vol.~8, pp. 913--916, 6
  2019.

\bibitem{8760267}
Z.~Wang, L.~Duan, and R.~Zhang, ``Adaptive deployment for uav-aided
  communication networks,'' \emph{IEEE Transactions on Wireless
  Communications}, vol.~18, no.~9, pp. 4531--4543, 2019.

\bibitem{Muntaha2021}
S.~T. Muntaha, S.~A. Hassan, H.~Jung, and M.~S. Hossain, ``Energy efficiency
  and hover time optimization in uav-based hetnets,'' \emph{IEEE Transactions
  on Intelligent Transportation Systems}, vol.~22, pp. 5103--5111, 8 2021.

\bibitem{9378782}
Q.-V. Pham, N.~Iradukunda, N.~H. Tran, W.-J. Hwang, and S.-H. Chung, ``Joint
  placement, power control, and spectrum allocation for uav wireless backhaul
  networks,'' \emph{IEEE Networking Letters}, vol.~3, no.~2, pp. 56--60, 2021.

\bibitem{8755983}
A.~Fouda, A.~S. Ibrahim, I.~Guvenc, and M.~Ghosh, ``Interference management in
  uav-assisted integrated access and backhaul cellular networks,'' \emph{IEEE
  Access}, vol.~7, pp. 104\,553--104\,566, 2019.

\bibitem{8771139}
A.~Betzler, D.~Camps-Mur, E.~Garcia-Villegas, I.~Demirkol, and J.~J.
  Aleixendri, ``Sodalite: Sdn wireless backhauling for dense 4g/5g small cell
  networks,'' \emph{IEEE Transactions on Network and Service Management},
  vol.~16, no.~4, pp. 1709--1723, 2019.

\bibitem{9423546}
M.~K.~Shehzad, A.~Ahmad, S.~A. Hassan, and H.~Jung, ``Backhaul-aware
  intelligent positioning of uavs and association of terrestrial base stations
  for fronthaul connectivity,'' \emph{IEEE Transactions on Network Science and
  Engineering}, vol.~8, no.~4, pp. 2742--2755, 2021.

\bibitem{9515712}
M.~Nikooroo and Z.~Becvar, ``Optimal positioning of flying base stations and
  transmission power allocation in noma networks,'' \emph{IEEE Transactions on
  Wireless Communications}, pp. 1--1, 2021.

\bibitem{9296771}
E.~H.~S. Cardoso, J.~P.~L. De~Araujo, S.~V. De~Carvalho, N.~Vijaykumar, and
  C.~R.~L. Frances, ``Novel multilayered cellular automata for flying cells
  positioning on 5g cellular self-organising networks,'' \emph{IEEE Access},
  vol.~8, pp. 227\,076--227\,099, 2020.

\bibitem{9373692}
R.~Zhang, R.~Lu, X.~Cheng, N.~Wang, and L.~Yang, ``A uav-enabled data
  dissemination protocol with proactive caching and file sharing in v2x
  networks,'' \emph{IEEE Transactions on Communications}, vol.~69, no.~6, pp.
  3930--3942, 2021.

\bibitem{9086619}
J.~Guo, P.~Walk, and H.~Jafarkhani, ``Optimal deployments of uavs with
  directional antennas for a power-efficient coverage,'' \emph{IEEE
  Transactions on Communications}, vol.~68, no.~8, pp. 5159--5174, 2020.

\bibitem{9044857}
S.-C. Noh, H.-B. Jeon, and C.-B. Chae, ``Energy-efficient deployment of
  multiple uavs using ellipse clustering to establish base stations,''
  \emph{IEEE Wireless Communications Letters}, vol.~9, no.~8, pp. 1155--1159,
  2020.

\bibitem{9293315}
W.~Feng, N.~Zhao, S.~Ao, J.~Tang, X.~Zhang, Y.~Fu, D.~K.~C. So, and K.-K. Wong,
  ``Joint 3d trajectory and power optimization for uav-aided mmwave mimo-noma
  networks,'' \emph{IEEE Transactions on Communications}, vol.~69, no.~4, pp.
  2346--2358, 2021.

\bibitem{9177323}
S.~T. Muntaha, S.~A. Hassan, H.~Jung, and M.~S. Hossain, ``Energy efficiency
  and hover time optimization in uav-based hetnets,'' \emph{IEEE Transactions
  on Intelligent Transportation Systems}, vol.~22, no.~8, pp. 5103--5111, 2021.

\bibitem{8320772}
A.~Fotouhi, M.~Ding, and M.~Hassan, ``Flying drone base stations for macro
  hotspots,'' \emph{IEEE Access}, vol.~6, pp. 19\,530--19\,539, 2018.

\bibitem{Zhang2021}
C.~Zhang, L.~Zhang, L.~Zhu, T.~Zhangb, Z.~Xiao, and X.-G. Xia, ``3d deployment
  of multiple uav-mounted base stations for uav communications,'' \emph{IEEE
  Transactions on Communications}, 2021.

\bibitem{Chen2020}
E.~Chen, J.~Chen, A.~W. Mohamed, B.~Wang, Z.~Wang, and Y.~Chen, ``Swarm
  intelligence application to uav aided iot data acquisition deployment
  optimization,'' \emph{IEEE Access}, vol.~8, pp. 175\,660--175\,668, 2020.

\bibitem{Hong2021}
D.~Hong, S.~Lee, Y.~H. Cho, D.~Baek, J.~Kim, and N.~Chang, ``Energy-efficient
  online path planning of multiple drones using reinforcement learning,''
  \emph{IEEE Transactions on Vehicular Technology}, vol.~70, pp. 9725--9740, 10
  2021.

\bibitem{Zhang12021}
K.~Zhang, X.~Gui, D.~Ren, and D.~Li, ``Energy-latency tradeoff for computation
  offloading in uav-assisted multiaccess edge computing system,'' \emph{IEEE
  Internet of Things Journal}, vol.~8, pp. 6709--6719, 4 2021.

\bibitem{Zhang2018}
S.~Zhang, H.~Zhang, B.~Di, and L.~Song, ``Joint trajectory and power
  optimization for uav sensing over cellular networks,'' \emph{IEEE
  Communications Letters}, vol.~22, pp. 2382--2385, 2018.

\bibitem{Hu2018}
J.~Hu, H.~Zhang, and L.~Song, ``Reinforcement learning for decentralized
  trajectory design in cellular uav networks with sense-and-send protocol,''
  \emph{IEEE Internet of Things Journal}, vol.~6, pp. 6177--6189, 2018.

\bibitem{Hua2020}
M.~Hua, L.~Yang, Q.~Wu, and A.~L. Swindlehurst, ``3d uav trajectory and
  communication design for simultaneous uplink and downlink transmission,''
  \emph{IEEE Transactions on Communications}, vol.~68, pp. 5908--5923, 2020.

\bibitem{Hiraguri2020}
T.~Hiraguri, K.~Nishimori, I.~Shitara, T.~Mitsui, T.~Shindo, T.~Kimura,
  T.~Matsuda, and H.~Yoshino, ``A cooperative transmission scheme in
  drone-based networks,'' \emph{IEEE Transactions on Vehicular Technology},
  vol.~69, pp. 2905--2914, 3 2020.

\bibitem{Wu2018}
Q.~Wu, Y.~Zeng, and R.~Zhang, ``Joint trajectory and communication design for
  multi-uav enabled wireless networks,'' \emph{IEEE Transactions on Wireless
  Communications}, vol.~17, pp. 2109--2121, 2018.

\bibitem{Ji2020}
J.~Ji, K.~Zhu, D.~Niyato, and R.~Wang, ``Joint cache placement, flight
  trajectory and transmission power optimization for multi-uav assisted
  wireless networks,'' \emph{IEEE Transactions on Wireless Communications},
  2020.

\bibitem{Tang2021}
Y.~Tang, Y.~Miao, A.~Barnawi, B.~Alzahrani, R.~Alotaibi, and K.~Hwang, ``A
  joint global and local path planning optimization for uav task scheduling
  towards crowd air monitoring,'' \emph{Computer Networks}, p. 107913, 2021.

\bibitem{Zhao2019}
N.~Zhao, X.~Pang, Z.~Li, Y.~Chen, F.~Li, Z.~Ding, and M.~S. Alouini, ``Joint
  trajectory and precoding optimization for uav-assisted noma networks,''
  \emph{IEEE Transactions on Communications}, vol.~67, pp. 3723--3735, 2019.

\bibitem{bertsimas-LPbook}
D.~Bertsimas and J.~Tsitsiklis, \emph{Introduction to linear
  optimization}.\hskip 1em plus 0.5em minus 0.4em\relax Athena Scientific,
  1997.

\end{thebibliography}
\end{document}